%
%
%
%
%
%
%
%
\def\standardrisposta{s }\def\reducedrisposta{r }
\def\mplarisposta{mpla }\def\zerorisposta{z }
\def\doublerisposta{d }\def\cartarisposta{e }\def\amsrisposta{y }
\newcount\ingrandimento \newcount\sinnota \newcount\dimnota
\newcount\unoduecol \newdimen\collhsize \newdimen\tothsize
\newdimen\fullhsize \newcount\controllorisposta \sinnota=1
\newskip\infralinea  \global\controllorisposta=0
\immediate\write16 { ********  Welcome to PANDA macros (Plain TeX,
AP, 1991) ******** }
%
%
%
%
%
%
\def\risposta{s } 
\def\srisposta{e } 
\def\arisposta{y }
\ifx\risposta\standardrisposta \ingrandimento=1200
\message {>> This will come out UNREDUCED << }
\dimnota=2 \unoduecol=1 \global\controllorisposta=1 \fi
\ifx\risposta\reducedrisposta \ingrandimento=1095 \dimnota=1
\unoduecol=1  \global\controllorisposta=1
\message {>> This will come out REDUCED << } \fi
\ifx\risposta\doublerisposta \ingrandimento=1000 \dimnota=2
\unoduecol=2

\message {>> You must print this in
LANDSCAPE orientation << } \global\controllorisposta=1 \fi
\ifx\risposta\mplarisposta \ingrandimento=1000 \dimnota=1
\message {>> Mod. Phys. Lett. A format << }
\unoduecol=1 \global\controllorisposta=1 \fi
\ifx\risposta\zerorisposta \ingrandimento=1000 \dimnota=2
\message {>> Zero Magnification format << }
\unoduecol=1 \global\controllorisposta=1 \fi
\ifnum\controllorisposta=0  \ingrandimento=1200
\message {>>> ERROR IN INPUT, I ASSUME STANDARD
UNREDUCED FORMAT <<< }  \dimnota=2 \unoduecol=1 \fi
\magnification=\ingrandimento
%
%
%
%
\newdimen\eucolumnsize \newdimen\eudoublehsize \newdimen\eudoublevsize
\newdimen\uscolumnsize \newdimen\usdoublehsize \newdimen\usdoublevsize
\newdimen\eusinglehsize \newdimen\eusinglevsize \newdimen\ussinglehsize
\newskip\standardbaselineskip \newdimen\ussinglevsize
\newskip\reducedbaselineskip \newskip\doublebaselineskip
\eucolumnsize=12.0truecm    
\eudoublehsize=25.5truecm   
\eudoublevsize=6.7truein    
\uscolumnsize=4.4truein     
\usdoublehsize=9.4truein    
\usdoublevsize=6.8truein    
\eusinglehsize=6.5truein    
\eusinglevsize=24truecm     
\ussinglehsize=6.5truein    
\ussinglevsize=8.9truein    
\standardbaselineskip=16pt plus.2pt  
\reducedbaselineskip=14pt plus.2pt   
\doublebaselineskip=12pt plus.2pt    
%
%
\def\Portoffset{}
\def\Landoffset{\voffset=-.2truein}
\ifx\risposta\mplarisposta \def\Portoffset{\hoffset=1.8truecm} \fi
%
%
\def\Landspec{}
\tolerance=10000
\parskip=0pt plus2pt  \leftskip=0pt \rightskip=0pt
%
%
\ifx\risposta\standardrisposta \infralinea=\standardbaselineskip \fi
\ifx\risposta\reducedrisposta  \infralinea=\reducedbaselineskip \fi
\ifx\risposta\doublerisposta   \infralinea=\doublebaselineskip \fi
\ifx\risposta\mplarisposta     \infralinea=13pt \fi
\ifx\risposta\zerorisposta     \infralinea=12pt plus.2pt\fi
\ifnum\controllorisposta=0    \infralinea=\standardbaselineskip \fi
\ifx\risposta\doublerisposta   \Landoffset \else \Portoffset \fi
\ifx\risposta\doublerisposta \ifx\srisposta\cartarisposta
\tothsize=\eudoublehsize \collhsize=\eucolumnsize
\vsize=\eudoublevsize  \else  \tothsize=\usdoublehsize
\collhsize=\uscolumnsize \vsize=\usdoublevsize \fi \else
\ifx\srisposta\cartarisposta \tothsize=\eusinglehsize
\vsize=\eusinglevsize \else  \tothsize=\ussinglehsize
\vsize=\ussinglevsize \fi \collhsize=4.4truein \fi
\ifx\risposta\mplarisposta \tothsize=5.0truein
\vsize=7.8truein \collhsize=4.4truein \fi
%
%
%
%
\newcount\contaeuler \newcount\contacyrill \newcount\contaams
\font\ninerm=cmr9  \font\eightrm=cmr8  \font\sixrm=cmr6
\font\ninei=cmmi9  \font\eighti=cmmi8  \font\sixi=cmmi6
\font\ninesy=cmsy9  \font\eightsy=cmsy8  \font\sixsy=cmsy6
\font\ninebf=cmbx9  \font\eightbf=cmbx8  \font\sixbf=cmbx6
\font\ninett=cmtt9  \font\eighttt=cmtt8  \font\nineit=cmti9
\font\eightit=cmti8 \font\ninesl=cmsl9  \font\eightsl=cmsl8
\skewchar\ninei='177 \skewchar\eighti='177 \skewchar\sixi='177
\skewchar\ninesy='60 \skewchar\eightsy='60 \skewchar\sixsy='60
\hyphenchar\ninett=-1 \hyphenchar\eighttt=-1 \hyphenchar\tentt=-1
\def\bfmath{\cmmib}                 
\font\tencmmib=cmmib10  \newfam\cmmibfam  \skewchar\tencmmib='177
\font\tencmbsy=cmbsy10  \newfam\cmbsyfam  \skewchar\tencmbsy='60
\def\scaps{\cmcsc}                 
\font\tencmcsc=cmcsc10  \newfam\cmcscfam
\ifnum\ingrandimento=1095

\font\capsone=cmcsc10 at 10.95pt \font\capstwo=cmcsc10 at 13.145pt

\else

\font\capsone=cmcsc10 at 12pt \font\capstwo=cmcsc10 at 14.4pt
\fi

\def\ttaarr{\bf}		
\def\ppaarr{\sl}		

%
%
%
\newfam\eufmfam \newfam\msamfam \newfam\msbmfam \newfam\eufbfam
\def\Loadeulerfonts{\global\contaeuler=1 \ifx\arisposta\amsrisposta
\font\teneufm=eufm10              
\font\eighteufm=eufm8 \font\nineeufm=eufm9 \font\sixeufm=eufm6
\font\seveneufm=eufm7  \font\fiveeufm=eufm5
\font\teneufb=eufb10              
\font\eighteufb=eufb8 \font\nineeufb=eufb9 \font\sixeufb=eufb6
\font\seveneufb=eufb7  \font\fiveeufb=eufb5
\font\teneurm=eurm10              
\font\eighteurm=eurm8 \font\nineeurm=eurm9
\font\teneurb=eurb10              
\font\eighteurb=eurb8 \font\nineeurb=eurb9
\font\teneusm=eusm10              
\font\eighteusm=eusm8 \font\nineeusm=eusm9
\font\teneusb=eusb10              
\font\eighteusb=eusb8 \font\nineeusb=eusb9
\else \def\eufm{\tt} \def\eufb{\tt} \def\eurm{\tt} \def\eurb{\tt}
\def\eusm{\tt} \def\eusb{\tt}    \fi}
\def\loadeuler{\Loadeulerfonts\tenpoint}
\def\loadamsmath{\global\contaams=1 \ifx\arisposta\amsrisposta
\font\tenmsam=msam10 \font\ninemsam=msam9 \font\eightmsam=msam8
\font\sevenmsam=msam7 \font\sixmsam=msam6 \font\fivemsam=msam5
\font\tenmsbm=msbm10 \font\ninemsbm=msbm9 \font\eightmsbm=msbm8
\font\sevenmsbm=msbm7 \font\sixmsbm=msbm6 \font\fivemsbm=msbm5
\else \def\msbm{\bf} \fi \def\Bbb{\msbm} \def\symbl{\msam} \tenpoint}
\def\loadcyrill{\global\contacyrill=1 \ifx\arisposta\amsrisposta
\font\tenwncyr=wncyr10 \font\ninewncyr=wncyr9 \font\eightwncyr=wncyr8
\font\tenwncyb=wncyr10 \font\ninewncyb=wncyr9 \font\eightwncyb=wncyr8
\font\tenwncyi=wncyr10 \font\ninewncyi=wncyr9 \font\eightwncyi=wncyr8
\else \def\cyrill{\sl} \def\cyrilb{\sl} \def\cyrili{\sl} \fi\tenpoint}
\ifx\arisposta\amsrisposta
\font\sevenex=cmex7               
\font\eightex=cmex8  \font\nineex=cmex9
\font\ninecmmib=cmmib9   \font\eightcmmib=cmmib8
\font\sevencmmib=cmmib7 \font\sixcmmib=cmmib6
\font\fivecmmib=cmmib5   \skewchar\ninecmmib='177
\skewchar\eightcmmib='177  \skewchar\sevencmmib='177
\skewchar\sixcmmib='177   \skewchar\fivecmmib='177
\font\ninecmbsy=cmbsy9    \font\eightcmbsy=cmbsy8
\font\sevencmbsy=cmbsy7  \font\sixcmbsy=cmbsy6
\font\fivecmbsy=cmbsy5   \skewchar\ninecmbsy='60
\skewchar\eightcmbsy='60  \skewchar\sevencmbsy='60
\skewchar\sixcmbsy='60    \skewchar\fivecmbsy='60
\font\ninecmcsc=cmcsc9    \font\eightcmcsc=cmcsc8     \else
\def\cmmib{\fam\cmmibfam\tencmmib}\textfont\cmmibfam=\tencmmib
\scriptfont\cmmibfam=\tencmmib \scriptscriptfont\cmmibfam=\tencmmib
\def\cmbsy{\fam\cmbsyfam\tencmbsy} \textfont\cmbsyfam=\tencmbsy
\scriptfont\cmbsyfam=\tencmbsy \scriptscriptfont\cmbsyfam=\tencmbsy
\scriptfont\cmcscfam=\tencmcsc \scriptscriptfont\cmcscfam=\tencmcsc
\def\cmcsc{\fam\cmcscfam\tencmcsc} \textfont\cmcscfam=\tencmcsc \fi
\catcode`@=11
\newskip\ttglue
\gdef\tenpoint{\def\rm{\fam0\tenrm}
  \textfont0=\tenrm \scriptfont0=\sevenrm \scriptscriptfont0=\fiverm
  \textfont1=\teni \scriptfont1=\seveni \scriptscriptfont1=\fivei
  \textfont2=\tensy \scriptfont2=\sevensy \scriptscriptfont2=\fivesy
  \textfont3=\tenex \scriptfont3=\tenex \scriptscriptfont3=\tenex
  \def\mcal{\fam2 \tensy}  \def\mmit{\fam1 \teni}
  \textfont\itfam=\tenit \def\it{\fam\itfam\tenit}
  \textfont\slfam=\tensl \def\sl{\fam\slfam\tensl}
  \textfont\ttfam=\tentt \scriptfont\ttfam=\eighttt
  \scriptscriptfont\ttfam=\eighttt  \def\tt{\fam\ttfam\tentt}
  \textfont\bffam=\tenbf \scriptfont\bffam=\sevenbf
  \scriptscriptfont\bffam=\fivebf \def\bf{\fam\bffam\tenbf}
     \ifx\arisposta\amsrisposta    \ifnum\contaeuler=1
  \textfont\eufmfam=\teneufm \scriptfont\eufmfam=\seveneufm
  \scriptscriptfont\eufmfam=\fiveeufm \def\eufm{\fam\eufmfam\teneufm}
  \textfont\eufbfam=\teneufb \scriptfont\eufbfam=\seveneufb
  \scriptscriptfont\eufbfam=\fiveeufb \def\eufb{\fam\eufbfam\teneufb}
  \def\eurm{\teneurm} \def\eurb{\teneurb} \def\eusm{\teneusm}
  \def\eusb{\teneusb}    \fi    \ifnum\contaams=1
  \textfont\msamfam=\tenmsam \scriptfont\msamfam=\sevenmsam
  \scriptscriptfont\msamfam=\fivemsam \def\msam{\fam\msamfam\tenmsam}
  \textfont\msbmfam=\tenmsbm \scriptfont\msbmfam=\sevenmsbm
  \scriptscriptfont\msbmfam=\fivemsbm \def\msbm{\fam\msbmfam\tenmsbm}
     \fi      \ifnum\contacyrill=1     \def\cyrill{\tenwncyr}
  \def\cyrilb{\tenwncyb}  \def\cyrili{\tenwncyi}         \fi
  \textfont3=\tenex \scriptfont3=\sevenex \scriptscriptfont3=\sevenex
  \def\cmmib{\fam\cmmibfam\tencmmib} \scriptfont\cmmibfam=\sevencmmib
  \textfont\cmmibfam=\tencmmib  \scriptscriptfont\cmmibfam=\fivecmmib
  \def\cmbsy{\fam\cmbsyfam\tencmbsy} \scriptfont\cmbsyfam=\sevencmbsy
  \textfont\cmbsyfam=\tencmbsy  \scriptscriptfont\cmbsyfam=\fivecmbsy
  \def\cmcsc{\fam\cmcscfam\tencmcsc} \scriptfont\cmcscfam=\eightcmcsc
  \textfont\cmcscfam=\tencmcsc \scriptscriptfont\cmcscfam=\eightcmcsc
     \fi            \tt \ttglue=.5em plus.25em minus.15em
  \normalbaselineskip=12pt
  \setbox\strutbox=\hbox{\vrule height8.5pt depth3.5pt width0pt}
  \let\sc=\eightrm \let\big=\tenbig   \normalbaselines
  \baselineskip=\infralinea  \rm}
\gdef\ninepoint{\def\rm{\fam0\ninerm}
  \textfont0=\ninerm \scriptfont0=\sixrm \scriptscriptfont0=\fiverm
  \textfont1=\ninei \scriptfont1=\sixi \scriptscriptfont1=\fivei
  \textfont2=\ninesy \scriptfont2=\sixsy \scriptscriptfont2=\fivesy
  \textfont3=\tenex \scriptfont3=\tenex \scriptscriptfont3=\tenex
  \def\mcal{\fam2 \ninesy}  \def\mmit{\fam1 \ninei}
  \textfont\itfam=\nineit \def\it{\fam\itfam\nineit}
  \textfont\slfam=\ninesl \def\sl{\fam\slfam\ninesl}
  \textfont\ttfam=\ninett \scriptfont\ttfam=\eighttt
  \scriptscriptfont\ttfam=\eighttt \def\tt{\fam\ttfam\ninett}
  \textfont\bffam=\ninebf \scriptfont\bffam=\sixbf
  \scriptscriptfont\bffam=\fivebf \def\bf{\fam\bffam\ninebf}
     \ifx\arisposta\amsrisposta  \ifnum\contaeuler=1
  \textfont\eufmfam=\nineeufm \scriptfont\eufmfam=\sixeufm
  \scriptscriptfont\eufmfam=\fiveeufm \def\eufm{\fam\eufmfam\nineeufm}
  \textfont\eufbfam=\nineeufb \scriptfont\eufbfam=\sixeufb
  \scriptscriptfont\eufbfam=\fiveeufb \def\eufb{\fam\eufbfam\nineeufb}
  \def\eurm{\nineeurm} \def\eurb{\nineeurb} \def\eusm{\nineeusm}
  \def\eusb{\nineeusb}     \fi   \ifnum\contaams=1
  \textfont\msamfam=\ninemsam \scriptfont\msamfam=\sixmsam
  \scriptscriptfont\msamfam=\fivemsam \def\msam{\fam\msamfam\ninemsam}
  \textfont\msbmfam=\ninemsbm \scriptfont\msbmfam=\sixmsbm
  \scriptscriptfont\msbmfam=\fivemsbm \def\msbm{\fam\msbmfam\ninemsbm}
     \fi       \ifnum\contacyrill=1     \def\cyrill{\ninewncyr}
  \def\cyrilb{\ninewncyb}  \def\cyrili{\ninewncyi}         \fi
  \textfont3=\nineex \scriptfont3=\sevenex \scriptscriptfont3=\sevenex
  \def\cmmib{\fam\cmmibfam\ninecmmib}  \textfont\cmmibfam=\ninecmmib
  \scriptfont\cmmibfam=\sixcmmib \scriptscriptfont\cmmibfam=\fivecmmib
  \def\cmbsy{\fam\cmbsyfam\ninecmbsy}  \textfont\cmbsyfam=\ninecmbsy
  \scriptfont\cmbsyfam=\sixcmbsy \scriptscriptfont\cmbsyfam=\fivecmbsy
  \def\cmcsc{\fam\cmcscfam\ninecmcsc} \scriptfont\cmcscfam=\eightcmcsc
  \textfont\cmcscfam=\ninecmcsc \scriptscriptfont\cmcscfam=\eightcmcsc
     \fi            \tt \ttglue=.5em plus.25em minus.15em
  \normalbaselineskip=11pt
  \setbox\strutbox=\hbox{\vrule height8pt depth3pt width0pt}
  \let\sc=\sevenrm \let\big=\ninebig \normalbaselines\rm}
\gdef\eightpoint{\def\rm{\fam0\eightrm}
  \textfont0=\eightrm \scriptfont0=\sixrm \scriptscriptfont0=\fiverm
  \textfont1=\eighti \scriptfont1=\sixi \scriptscriptfont1=\fivei
  \textfont2=\eightsy \scriptfont2=\sixsy \scriptscriptfont2=\fivesy
  \textfont3=\tenex \scriptfont3=\tenex \scriptscriptfont3=\tenex
  \def\mcal{\fam2 \eightsy}  \def\mmit{\fam1 \eighti}
  \textfont\itfam=\eightit \def\it{\fam\itfam\eightit}
  \textfont\slfam=\eightsl \def\sl{\fam\slfam\eightsl}
  \textfont\ttfam=\eighttt \scriptfont\ttfam=\eighttt
  \scriptscriptfont\ttfam=\eighttt \def\tt{\fam\ttfam\eighttt}
  \textfont\bffam=\eightbf \scriptfont\bffam=\sixbf
  \scriptscriptfont\bffam=\fivebf \def\bf{\fam\bffam\eightbf}
     \ifx\arisposta\amsrisposta   \ifnum\contaeuler=1
  \textfont\eufmfam=\eighteufm \scriptfont\eufmfam=\sixeufm
  \scriptscriptfont\eufmfam=\fiveeufm \def\eufm{\fam\eufmfam\eighteufm}
  \textfont\eufbfam=\eighteufb \scriptfont\eufbfam=\sixeufb
  \scriptscriptfont\eufbfam=\fiveeufb \def\eufb{\fam\eufbfam\eighteufb}
  \def\eurm{\eighteurm} \def\eurb{\eighteurb} \def\eusm{\eighteusm}
  \def\eusb{\eighteusb}       \fi    \ifnum\contaams=1
  \textfont\msamfam=\eightmsam \scriptfont\msamfam=\sixmsam
  \scriptscriptfont\msamfam=\fivemsam \def\msam{\fam\msamfam\eightmsam}
  \textfont\msbmfam=\eightmsbm \scriptfont\msbmfam=\sixmsbm
  \scriptscriptfont\msbmfam=\fivemsbm \def\msbm{\fam\msbmfam\eightmsbm}
     \fi       \ifnum\contacyrill=1     \def\cyrill{\eightwncyr}
  \def\cyrilb{\eightwncyb}  \def\cyrili{\eightwncyi}         \fi
  \textfont3=\eightex \scriptfont3=\sevenex \scriptscriptfont3=\sevenex
  \def\cmmib{\fam\cmmibfam\eightcmmib}  \textfont\cmmibfam=\eightcmmib
  \scriptfont\cmmibfam=\sixcmmib \scriptscriptfont\cmmibfam=\fivecmmib
  \def\cmbsy{\fam\cmbsyfam\eightcmbsy}  \textfont\cmbsyfam=\eightcmbsy
  \scriptfont\cmbsyfam=\sixcmbsy \scriptscriptfont\cmbsyfam=\fivecmbsy
  \def\cmcsc{\fam\cmcscfam\eightcmcsc} \scriptfont\cmcscfam=\eightcmcsc
  \textfont\cmcscfam=\eightcmcsc \scriptscriptfont\cmcscfam=\eightcmcsc
     \fi             \tt \ttglue=.5em plus.25em minus.15em
  \normalbaselineskip=9pt
  \setbox\strutbox=\hbox{\vrule height7pt depth2pt width0pt}
  \let\sc=\sixrm \let\big=\eightbig \normalbaselines\rm }
\gdef\tenbig#1{{\hbox{$\left#1\vbox to8.5pt{}\right.\n@space$}}}
\gdef\ninebig#1{{\hbox{$\textfont0=\tenrm\textfont2=\tensy
   \left#1\vbox to7.25pt{}\right.\n@space$}}}
\gdef\eightbig#1{{\hbox{$\textfont0=\ninerm\textfont2=\ninesy
   \left#1\vbox to6.5pt{}\right.\n@space$}}}
\def\alternativefont#1#2{\ifx\arisposta\amsrisposta \relax \else
\xdef#1{#2} \fi}
\global\contaeuler=0 \global\contacyrill=0 \global\contaams=0
%
%
%
%
\newbox\fotlinebb \newbox\hedlinebb \newbox\leftcolumn
\gdef\makeheadline{\vbox to 0pt{\vskip-22.5pt
     \fullline{\vbox to8.5pt{}\the\headline}\vss}\nointerlineskip}
\gdef\makehedlinebb{\vbox to 0pt{\vskip-22.5pt
     \fullline{\vbox to8.5pt{}\copy\hedlinebb\hfil
     \line{\hfill\the\headline\hfill}}\vss} \nointerlineskip}
\gdef\makefootline{\baselineskip=24pt \fullline{\the\footline}}
\gdef\makefotlinebb{\baselineskip=24pt
    \fullline{\copy\fotlinebb\hfil\line{\hfill\the\footline\hfill}}}
\gdef\doubleformat{\shipout\vbox{\Landspec\makehedlinebb
     \fullline{\box\leftcolumn\hfil\columnbox}\makefotlinebb}
     \advancepageno}
\gdef\columnbox{\leftline{\pagebody}}
\gdef\line#1{\hbox to\hsize{\hskip\leftskip#1\hskip\rightskip}}
\gdef\fullline#1{\hbox to\fullhsize{\hskip\leftskip{#1}%
\hskip\rightskip}}
\gdef\footnote#1{\let\@sf=\empty
         \ifhmode\edef\#sf{\spacefactor=\the\spacefactor}\/\fi
         #1\@sf\vfootnote{#1}}
\gdef\vfootnote#1{\insert\footins\bgroup
         \ifnum\dimnota=1  \eightpoint\fi
         \ifnum\dimnota=2  \ninepoint\fi
         \ifnum\dimnota=0  \tenpoint\fi
         \interlinepenalty=\interfootnotelinepenalty
         \splittopskip=\ht\strutbox
         \splitmaxdepth=\dp\strutbox \floatingpenalty=20000
         \leftskip=\oldssposta \rightskip=\olddsposta
         \spaceskip=0pt \xspaceskip=0pt
         \ifnum\sinnota=0   \textindent{#1}\fi
         \ifnum\sinnota=1   \item{#1}\fi
         \footstrut\futurelet\next\fo@t}
\gdef\fo@t{\ifcat\bgroup\noexpand\next \let\next\f@@t
             \else\let\next\f@t\fi \next}
\gdef\f@@t{\bgroup\aftergroup\@foot\let\next}
\gdef\f@t#1{#1\@foot} \gdef\@foot{\strut\egroup}
\gdef\footstrut{\vbox to\splittopskip{}}
\skip\footins=\bigskipamount
\count\footins=1000  \dimen\footins=8in
\catcode`@=12
\tenpoint
\ifnum\unoduecol=1 \hsize=\tothsize   \fullhsize=\tothsize \fi
\ifnum\unoduecol=2 \hsize=\collhsize  \fullhsize=\tothsize \fi
\global\let\lrcol=L      \ifnum\unoduecol=1
\output{\plainoutput{\ifnum\tipbnota=2 \clearnmbnota\fi}} \fi
\ifnum\unoduecol=2 \output{\if L\lrcol
     \global\setbox\leftcolumn=\columnbox
     \global\setbox\fotlinebb=\line{\hfill\the\footline\hfill}
     \global\setbox\hedlinebb=\line{\hfill\the\headline\hfill}
     \advancepageno  \global\let\lrcol=R
     \else  \doubleformat \global\let\lrcol=L \fi
     \ifnum\outputpenalty>-20000 \else\dosupereject\fi
     \ifnum\tipbnota=2\clearnmbnota\fi }\fi
\def\ifdoublepage{\ifnum\unoduecol=2 }
\gdef\yespagenumbers{\footline={\hss\tenrm\folio\hss}}
\gdef\ciao{ \ifnum\fdefcontre=1 \endfdef\fi
     \par\vfill\supereject \ifnum\unoduecol=2
     \if R\lrcol  \headline={}\nopagenumbers\null\vfill\eject
     \fi\fi \end}

\newskip\olddsposta \newskip\oldssposta
\global\oldssposta=\leftskip \global\olddsposta=\rightskip

\def\filldots{\leaders\hbox to 1em{\hss.\hss}\hfill}
\def\inquadrb#1 {\vbox {\hrule  \hbox{\vrule \vbox {\vskip .2cm
    \hbox {\ #1\ } \vskip .2cm } \vrule  }  \hrule} }
 \def\newline{\hfil\break}
\def\jump{\vskip\baselineskip} \newskip\iinnffrr
\def\sjump{\iinnffrr=\baselineskip
          \divide\iinnffrr by 2 \vskip\iinnffrr}
\def\bjump{\vskip\baselineskip \vskip\baselineskip}
\newcount\nmbnota  \def\clearnmbnota{\global\nmbnota=0}
\newcount\tipbnota \def\letterfootnote{\global\tipbnota=1}

\def\note#1{\global\advance\nmbnota by 1 \ifnum\tipbnota=1
    \footnote{$^{\rm\nttlett}$}{#1} \else {\ifnum\tipbnota=2
    \footnote{$^{\nttsymb}$}{#1}
    \else\footnote{$^{\the\nmbnota}$}{#1}\fi}\fi}
\def\nttlett{\ifcase\nmbnota \or a\or b\or c\or d\or e\or f\or
g\or h\or i\or j\or k\or l\or m\or n\or o\or p\or q\or r\or
s\or t\or u\or v\or w\or y\or x\or z\fi}
\def\nttsymb{\ifcase\nmbnota \or\dag\or\sharp\or\ddag\or\star\or
\natural\or\flat\or\clubsuit\or\diamondsuit\or\heartsuit
\or\spadesuit\fi}   \clearnmbnota
\def\numberfootnote{\global\tipbnota=0} \numberfootnote
\def\setnote#1{\expandafter\xdef\csname#1\endcsname{
\ifnum\tipbnota=1 {\rm\nttlett} \else {\ifnum\tipbnota=2
{\nttsymb} \else \the\nmbnota\fi}\fi} }
\newcount\nbmfig  \def\clearnbmfig{\global\nbmfig=0}
\gdef\figure{\global\advance\nbmfig by 1
      {\rm fig. \the\nbmfig}}   \clearnbmfig
\def\setfig#1{\expandafter\xdef\csname#1\endcsname{fig. \the\nbmfig}}
 \def\endformula{\eqno\numero $$}
 \def\efr{\endformula}
\newcount\frmcount \def\clearfrmcount{\global\frmcount=0}
\def\numero{\global\advance\frmcount by 1   \ifnum\indappcount=0
  {\ifnum\cpcount <1 {\hbox{\rm (\the\frmcount )}}  \else
  {\hbox{\rm (\the\cpcount .\the\frmcount )}} \fi}  \else
  {\hbox{\rm (\applett .\the\frmcount )}} \fi}
\def\nameformula#1{\global\advance\frmcount by 1%
\ifnum\draftnum=0  {\ifnum\indappcount=0%
{\ifnum\cpcount<1\xdef\spzzttrra{(\the\frmcount )}%
\else\xdef\spzzttrra{(\the\cpcount .\the\frmcount )}\fi}%
\else\xdef\spzzttrra{(\applett .\the\frmcount )}\fi}%
\else\xdef\spzzttrra{(#1)}\fi%
\expandafter\xdef\csname#1\endcsname{\spzzttrra}
\eqno \hbox{\rm\spzzttrra} $$}
\def\nfr{\nameformula}    
\def\nameali#1{\global\advance\frmcount by 1%
\ifnum\draftnum=0  {\ifnum\indappcount=0%
{\ifnum\cpcount<1\xdef\spzzttrra{(\the\frmcount )}%
\else\xdef\spzzttrra{(\the\cpcount .\the\frmcount )}\fi}%
\else\xdef\spzzttrra{(\applett .\the\frmcount )}\fi}%
\else\xdef\spzzttrra{(#1)}\fi%
\expandafter\xdef\csname#1\endcsname{\spzzttrra}
  \hbox{\rm\spzzttrra} }      \clearfrmcount
\newcount\cpcount \def\clearcpcount{\global\cpcount=0}
\newcount\subcpcount \def\clearsubcpcount{\global\subcpcount=0}
\newcount\appcount \def\clearappcount{\global\appcount=0}
\newcount\indappcount \def\clearindappcount{\indappcount=0}
\newcount\sottoparcount 

\def\applett{\ifcase\appcount  \or {A}\or {B}\or {C}\or
{D}\or {E}\or {F}\or {G}\or {H}\or {I}\or {J}\or {K}\or {L}\or
{M}\or {N}\or {O}\or {P}\or {Q}\or {R}\or {S}\or {T}\or {U}\or
{V}\or {W}\or {X}\or {Y}\or {Z}\fi    \ifnum\appcount<0
\immediate\write16 {Panda ERROR - Appendix: counter "appcount"
out of range}\fi  \ifnum\appcount>26  \immediate\write16 {Panda
ERROR - Appendix: counter "appcount" out of range}\fi}
\clearappcount  \clearindappcount \newcount\connttrre
\def\clearconnttrre{\global\connttrre=0} \newcount\countref
\def\clearcountref{\global\countref=0} \clearcountref
\def\chapter#1{\global\advance\cpcount by 1 \clearfrmcount
                 \goodbreak\null\vbox{\jump\nobreak
                 \clearsubcpcount\clearindappcount
                 \itemitem{\ttaarr\the\cpcount .\qquad}{\ttaarr #1}
                 \par\nobreak\jump\sjump}\nobreak}
\def\section#1{\global\advance\subcpcount by 1 \goodbreak\null
               \vbox{\sjump\nobreak\ifnum\indappcount=0
                 {\ifnum\cpcount=0 {\itemitem{\ppaarr
               .\the\subcpcount\quad\enskip\ }{\ppaarr #1}\par} \else
                 {\itemitem{\ppaarr\the\cpcount .\the\subcpcount\quad
                  \enskip\ }{\ppaarr #1} \par}  \fi}
                \else{\itemitem{\ppaarr\applett .\the\subcpcount\quad
                 \enskip\ }{\ppaarr #1}\par}\fi\nobreak\jump}\nobreak}
\clearsubcpcount
\def\appendix#1{\global\advance\appcount by 1 \clearfrmcount
                  \goodbreak\null\vbox{\jump\nobreak
                  \global\advance\indappcount by 1 \clearsubcpcount
          \itemitem{ }{\hskip-40pt\ttaarr #1}
             \nobreak\jump\sjump}\nobreak}
\clearappcount \clearindappcount
\def\references{\goodbreak\null\vbox{\jump\nobreak
   \noindent{\ttaarr References} \nobreak\jump\sjump}\nobreak}

\clearcpcount\clearcountref

\def\setchap#1{\ifnum\indappcount=0{\ifnum\subcpcount=0%
\xdef\spzzttrra{\the\cpcount}%
\else\xdef\spzzttrra{\the\cpcount .\the\subcpcount}\fi}
\else{\ifnum\subcpcount=0 \xdef\spzzttrra{\applett}%
\else\xdef\spzzttrra{\applett .\the\subcpcount}\fi}\fi
\expandafter\xdef\csname#1\endcsname{\spzzttrra}}
\newcount\draftnum \newcount\ppora   \newcount\ppminuti
\global\ppora=\time   \global\ppminuti=\time
\global\divide\ppora by 60  \draftnum=\ppora
\multiply\draftnum by 60    \global\advance\ppminuti by -\draftnum
\def\droggi{\number\day /\number\month /\number\year\ \the\ppora
:\the\ppminuti}     \global\draftnum=0
\def\draftcomment#1{\ifnum\draftnum=0 \relax \else
{\ {\bf ***}\ #1\ {\bf ***}\ }\fi} 
%
%
\catcode`@=11
\gdef\Ref#1{\expandafter\ifx\csname @rrxx@#1\endcsname\relax%
{\global\advance\countref by 1    \ifnum\countref>200
\immediate\write16 {Panda ERROR - Ref: maximum number of references
exceeded}  \expandafter\xdef\csname @rrxx@#1\endcsname{0}\else
\expandafter\xdef\csname @rrxx@#1\endcsname{\the\countref}\fi}\fi
\ifnum\draftnum=0 \csname @rrxx@#1\endcsname \else#1\fi}
\gdef\beginref{\ifnum\draftnum=0  \gdef\Rref{\fairef}
\gdef\endref{\scriviref} \else\relax\fi
\ifx\risposta\mplarisposta \ninepoint \fi
\parskip 2pt plus.2pt \baselineskip=12pt}
\def\Reflab#1{[#1]} \gdef\Rref#1#2{\item{\Reflab{#1}}{#2}}
\gdef\endref{\relax}  \newcount\conttemp
\gdef\fairef#1#2{\expandafter\ifx\csname @rrxx@#1\endcsname\relax
{\global\conttemp=0 \immediate\write16 {Panda ERROR - Ref: reference
[#1] undefined}} \else
{\global\conttemp=\csname @rrxx@#1\endcsname } \fi
\global\advance\conttemp by 50  \global\setbox\conttemp=\hbox{#2} }
\gdef\scriviref{\clearconnttrre\conttemp=50
\loop\ifnum\connttrre<\countref \advance\conttemp by 1
\advance\connttrre by 1
\item{\Reflab{\the\connttrre}}{\unhcopy\conttemp} \repeat}
\clearcountref \clearconnttrre
\catcode`@=12
\ifx\risposta\mplarisposta \def\Reflab#1{#1.} \letterfootnote \fi

\def\slashchar#1{\setbox0=\hbox{$#1$} \dimen0=\wd0
     \setbox1=\hbox{/} \dimen1=\wd1 \ifdim\dimen0>\dimen1
      \rlap{\hbox to \dimen0{\hfil/\hfil}} #1 \else
      \rlap{\hbox to \dimen1{\hfil$#1$\hfil}} / \fi}
\ifx\oldchi\undefined \let\oldchi=\chi
  \def\cchi{{\raise 1pt\hbox{$\oldchi$}}} \let\chi=\cchi \fi

\def\frac#1#2{{\textstyle{#1 \over #2}}}

\def\half{\ifinner {\scriptstyle {1 \over 2}}\else {1 \over 2} \fi}
\def\bra#1{\langle#1\vert}  \def\ket#1{\vert#1\rangle}

\def\simge{\rlap{\raise 2pt \hbox{$>$}}{\lower 2pt \hbox{$\sim$}}}
\def\simle{\rlap{\raise 2pt \hbox{$<$}}{\lower 2pt \hbox{$\sim$}}}

\def\buildchar#1#2#3{{\null\!\mathop{#1}\limits^{#2}_{#3}\!\null}}

\def\vbig#1#2{{\vbigd@men=#2\divide\vbigd@men by 2%
\hbox{$\left#1\vbox to \vbigd@men{}\right.\n@space$}}}

%
%
\newcount\fdefcontre \newcount\fdefcount \newcount\indcount
\newread\filefdef  \newread\fileftmp  \newwrite\filefdef
\newwrite\fileftmp     \def\strip#1*.A {#1}
\def\futuredef#1{\beginfdef
\expandafter\ifx\csname#1\endcsname\relax%
{\immediate\write\fileftmp {#1*.A}
\immediate\write16 {Panda Warning - fdef: macro "#1" on page
\the\pageno \space undefined}
\ifnum\draftnum=0 \expandafter\xdef\csname#1\endcsname{(?)}
\else \expandafter\xdef\csname#1\endcsname{(#1)} \fi
\global\advance\fdefcount by 1}\fi   \csname#1\endcsname}

\def\beginfdef{\ifnum\fdefcontre=0
\immediate\openin\filefdef \jobname.fdef
\immediate\openout\fileftmp \jobname.ftmp
\global\fdefcontre=1  \ifeof\filefdef \immediate\write16 {Panda
WARNING - fdef: file \jobname.fdef not found, run TeX again}
\else \immediate\read\filefdef to\spzzttrra
\global\advance\fdefcount by \spzzttrra
\indcount=0      \loop\ifnum\indcount<\fdefcount
\advance\indcount by 1   \immediate\read\filefdef to\spezttrra
\immediate\read\filefdef to\sppzttrra
\edef\spzzttrra{\expandafter\strip\spezttrra}
\immediate\write\fileftmp {\spzzttrra *.A}
\expandafter\xdef\csname\spzzttrra\endcsname{\sppzttrra}
\repeat \fi \immediate\closein\filefdef \fi}
\def\endfdef{\immediate\closeout\fileftmp   \ifnum\fdefcount>0
\immediate\openin\fileftmp \jobname.ftmp
\immediate\openout\filefdef \jobname.fdef
\immediate\write\filefdef {\the\fdefcount}   \indcount=0
\loop\ifnum\indcount<\fdefcount    \advance\indcount by 1
\immediate\read\fileftmp to\spezttrra
\edef\spzzttrra{\expandafter\strip\spezttrra}
\immediate\write\filefdef{\spzzttrra *.A}
\edef\spezttrra{\string{\csname\spzzttrra\endcsname\string}}
\iwritel\filefdef{\spezttrra}
\repeat  \immediate\closein\fileftmp \immediate\closeout\filefdef
\immediate\write16 {Panda Warning - fdef: Label(s) may have changed,
re-run TeX to get them right}\fi}
\def\iwritel#1#2{\newlinechar=-1
{\newlinechar=`\ \immediate\write#1{#2}}\newlinechar=-1}
\global\fdefcontre=0 \global\fdefcount=0 \global\indcount=0
%
%
\null
%
%
%
%
\loadamsmath
\loadeuler
\mathchardef\hbar="2D7D
\def\gg{{\>\widehat{g}\>}}
\def\sh{{\bf s}_{\rm h}}
\def\ss{{\bf s}}
\def\cc{{\bf c}}
\def\hb{{\bfmath h}}
\mathchardef\blamb="7315
\mathchardef\bLamb="7303
\mathchardef\bmu="7316
\mathchardef\balpha="710B
\mathchardef\bbeta="710C
\mathchardef\bgamma="710D
\def\lab{{\bfmath\blamb}}
\def\Lab{{\bfmath\bLamb}}
\def\mub{{\bfmath\bmu}}
\def\alb{{\bfmath\balpha}}
\def\beb{{\bfmath\bbeta}}
\def\gab{{\bfmath\bgamma}}
\pageno=0
\nopagenumbers{\baselineskip=12pt
\line{\hfill US-FT/19-97}
\line{\hfill\tt hep-th/9706203}
\line{\hfill June 1997}\bjump
\ifdoublepage \bjump\bjump\bjump\else\jump\vfill\fi
\centerline{\capstwo Semi-Classical Spectrum of the}
\jump
\centerline{\capstwo Homogeneous sine-Gordon theories}
\bjump\jump
\centerline{{\scaps Carlos R.~Fern\'andez-Pousa}~ and~ {\scaps J. Luis
Miramontes}}
\jump\jump
\centerline{\sl Departamento de F\'\i sica de Part\'\i
culas,}
\centerline{\sl Facultad de F\'\i sica,}
\centerline{\sl Universidad de Santiago,}
\centerline{\sl E-15706 Santiago de Compostela, Spain}
\jump
\centerline{{\tt pousa@gaes.usc.es~} and~ {\tt miramont@fpaxp1.usc.es}} 
\bjump\bjump
\ifdoublepage
\vfill {\noindent
\line{June 1997\hfill}}
\eject\null\vfill\fi
\centerline{\capsone ABSTRACT}\jump

\noindent
The semi-classical spectrum of the Homogeneous sine-Gordon theories associated
with an arbitrary compact simple Lie group~$G$ is obtained and shown to
be entirely given by solitons. These theories describe quantum integrable
massive perturbations of Gepner's $G$-parafermions whose classical
equations-of-motion are non-abelian affine Toda equations. One-soliton
solutions are constructed by embeddings of the $SU(2)$ complex sine-Gordon
soliton in the regular $SU(2)$ subgroups of $G$. The resulting spectrum
exhibits both stable and unstable particles, which is a peculiar feature
shared with the spectrum of monopoles and dyons in $N=2$ and $N=4$
supersymmetric gauge theories.

\vfill
\ifdoublepage \else
\noindent
\line{June 1997\hfill}\fi
\eject}
\yespagenumbers\pageno=1
\footline={\hss\tenrm-- \folio\ --\hss}

\chapter{Introduction}

The idea that extended particles in quantum field theory can be associated 
with soliton solutions of the classical equations-of-motion dates back to
the early work of Skyrme~[\Ref{SKYR}]. Well known examples of this are
provided by the Skyrme model, where solitons are identified with the baryons
of large-$N_c$ QCD~[\Ref{ADK}], and by monopoles and dyons, which arise as
solitons in non-abelian gauge theories~[\Ref{MODY}]. In most cases, the
spectrum and interactions of this kind of particles can only be found at weak
coupling limit, and it is extremely useful to have simple solvable models that
capture as many features of realistic solitons as possible. In this sense,
integrability has proved to be a powerful tool and, in fact, an important
paradigm for Skyrme's ideas is provided by the sine-Gordon (SG) field theory,
where an exact description of the soliton dynamics can be deduced~[\Ref{SG}],
which matches precisely with semi-classical approaches in the appropriate
limit. Results like this support our confidence that similar semi-classical
approaches are equally valid in more realistic theories in higher dimensions.

Many integrable generalizations of the SG equations-of-motion have been
written down and are known as the (abelian and non-abelian) affine Toda
equations. However, not all these equations can be understood as the
equations-of-motion of an action with sensible properties like a
positive-definite kinetic term and a real potential. Actually, it has been
shown in~[\Ref{MASS}] that these requirements, together with the existence of
soliton solutions, select only two series of affine Toda equations that give
rise to theories with a mass-gap and, hence, that are expected to admit an
$S$-matrix description. In that article, these two series of theories are
referred to as Homogeneous sine-Gordon (HSG) theories and Symmetric Space
sine-Gordon (SSSG) theories, which are associated with compact simple Lie
groups and compact symmetric spaces, respectively. 

The HSG and SSSG theories describe integrable perturbations of
$c>1$~conformal field theories which, in the classical limit, are formulated
by means of a unitary Lagrangian. In contrast, recall the case of the abelian
affine Toda field theories. There, both their interpretation as integrable
perturbations of the ($c<1$) minimal models~[\Ref{EYHM}] and the existence of
soliton solutions~[\Ref{SOLC}] require an imaginary coupling constant, which,
unless for the sine-Gordon theory, renders their potential
complex~[\Ref{JAPOS}]. An important consequence of having a proper unitary
Lagrangian description is the possibility of addressing the quantum properties
of the HSG and SSSG theories by using both the standard methods of quantum
field theory, and the techniques originally designed by Zamolodchikov for
studying integrability~[\Ref{ZAMINT}].
 
The purpose of this paper is to study the classical and semi-classical
soliton spectrum of the HSG theories, whose quantum integrability has been
established in~[\Ref{QNOS}]. The construction and main features of these
theories are briefly summarized in Section~2. Their classical action consists
of the gauged Wess-Zumino-Witten (WZW) action corresponding to the
$G/U(1)^{\times r_g}$ coset model perturbed by a potential, where $G$ is a
compact simple Lie group of rank~$r_g$. This action exhibits a global
$U(1)^{\times r_g}$ symmetry and, hence, solitons carry a conserved (Noether)
vector charge ${\bfmath Q}$. At the quantum level, the HSG theories describe
massive perturbations of the the theories of
$G$-parafermions~[\Ref{PARAF},\Ref{GPARAF}] where the potential induces the
complete breaking of the conformal symmetry. The simplest HSG theory,
associated with $G=SU(2)$, corresponds to the perturbation of the
${\Bbb Z}_k$-parafermions by the first thermal operator, and its
equation-of-motion is the complex sine-Gordon (CSG)
equation~[\Ref{BAK},\Ref{PARK1}].

Out of the different methods available to construct explicit soliton
solutions of non-abelian affine Toda equations, we will use the so-called
`solitonic specialization' proposed by Olive {\it et
al.\/}~[\Ref{OLIVET},\Ref{OLIVE}], whose relation with the method of dressing
transformations and $\tau$-functions has been recently clarified
in~[\Ref{TAUS}]. The application of this method to the HSG theories is
discussed in Section~3, and we use it in Section~4 to obtain the one-soliton
solutions of the theories associated with simply-laced Lie groups. In
this case, explicit solutions are obtained by means of the homogeneous vertex
operator construction of the untwisted affine Kac-Moody algebra associated
with~$g$. The main result of this section is that there is a one-soliton
solution associated with each root~$\alb$ of~$g$ that is nothing
else but a CSG soliton embedded in the regular $SU(2)$ subgroup of $G$
generated by $E_{\alb}$, $E_{-\alb}$, and $[E_{\alb}, E_{-\alb}]$.

This result is exploited in Section~5 to construct the one-soliton solutions
of the HSG theories associated with arbitrary compact simple Lie groups by
embedding the CSG soliton in the regular $SU(2)$ subgroups of $G$, which are
labelled by the roots of $g$. Actually, this method is widely used in the
context of Yang-Mills theories based on arbitrary Lie groups to construct
monopole or instanton solutions by embeddings of the $SU(2)$ spherically
symmetric 't-Hooft-Polyakov monopole~[\Ref{MONOS},\Ref{MONOSUSY}] or the
self-dual $SU(2)$ Belavin-Polyakov-Schwartz-Tyupkin instanton~[\Ref{INST}]. The
$SU(2)$ CSG soliton plays a similar building block role in the HSG theories.

In their rest frame, HSG solitons provide periodic time-dependent solutions
describing the rotation in the internal $U(1)^{\times r_g}$ space and, in
Section~6, we apply the Bohr-Sommerfeld quantization rule to obtain the
semi-classical spectrum. After quantization, the classical coupling constant
of the theory becomes the level~$k$ of the underlying $G$-parafermionic
theory, and the vector charge ${\bfmath Q}$ is restricted to take values in
the root lattice of $g$, $\Lab_R$, modulo $k$ times the co-root lattice,
$\Lab_{R}^\ast$, which is precisely the discrete symmetry group of the
Gepner theory of $G$-parafermions at level-$k$ ~[\Ref{GPARAF}].
For each root, there is a finite tower of massive particles whose number
depends on the length of the root. Unlike the SG kinks, CSG and HSG solitons
do not have topological quantum numbers and, as a consequence, the fundamental
particle associated with a root $\alb$ can be identified with the ${\bfmath
Q}=\alb$ soliton itself. Therefore, the full spectrum of the HSG theories is
described by solitons. 

One of the most novel features of solitons captured by the HSG theories is the
existence of unstable particles. In fact, when $r_g\geq 2$, only the soliton
particles associated with the simple roots are stable, while the other are
either unstable (resonances) or bound-states at threshold. An identical
phenomenon occurs for monopoles and dyons in $N=2$ and $N=4$ supersymmetric
gauge theories~[\Ref{MONOSUSY}], which provides extra motivation for the
overall aim of finding a factorizable $S$-matrix capable of describing the
quantum scattering of the solitons constructed in this paper.

Our conclusions are presented in Section~7 where we also comment on the
relation between solitons and parafermions. We have collected our conventions
about Kac-Moody algebras and vertex operators in an appendix.

\chapter{The Homogeneous sine-Gordon theories}

The construction of the HSG theory associated with a given compact simple
Lie group $G$ starts with the choice of two constant elements
$\Lambda_\pm$ in $g$, the Lie algebra of $G$. The condition
that the resulting theory has a mass-gap requires $\Lambda_\pm$ 
to be semisimple and regular, which means that their centralizer in
$g$ is a Cartan subalgebra. Otherwise, the choice of $\Lambda_\pm$ is
completely free and, hence, they can be considered as continuous
vector couplings of the theory. To comply with the notation 
of~[\Ref{MASS}], where more details can be found, we will refer to the
centralizer of $\Lambda_+$ in $g$ and to the corresponding abelian group as
$g_{0}^0 = u(1)^{+ r_g}$ and $G_{0}^0= U(1)^{\times r_g}$,
respectively, where $r_g$ is the  rank of $G$. The HSG theory is specified by
the action
$$
S[h, A_\pm]\>=\> {1\over \beta^2} \biggl\{ S_{\rm WZW}[h,A_\pm] 
\>-\>\int d^2 x \>V(h) \biggl\}\>.
\nfr{Act}
Here, $h$ is a bosonic field that takes values in $G$, $A_\pm$ are
(non-dymanical) abelian gauge connections taking values in $g_{0}^0$,
and $S_{\rm WZW}[h,A_\pm]$ is the gauged Wess-Zumino-Witten
action describing the $G/G_{0}^0 = G/U(1)^{\times r_g}$ coset model,  
$$
\eqalign{
S_{\rm WZW}[h & ,A_\pm]\> = \> S_{\rm WZW}[h]\> + \> {1\over \pi}  
\int d^2x \> \Bigl( -\langle A_+\>,\> \partial_-h\> h^{\dagger}\rangle\cr  
& + \> \langle \tau(A_-)\>, \>
h^{\dagger}\> \partial_+h\rangle\> +\> \langle h^{\dagger}\> A_+\> 
h\>, \> \tau(A_-) \rangle\> -\> \langle A_+\>,\> A_-\rangle\Bigr)\>,\cr}
\nfr{WZW}
where $x_\pm = t\pm x$ are light-cone variables in the $1+1$ 
Minkowski space. The potential
$V(h)$ is
$$
V(h)\> =\> - \> {m^2 \over \pi}\> \langle  \Lambda_+\>,\>  
h^{\dagger}
\>
\Lambda_- \> h\rangle\>,
\nfr{Pot}
and we have denoted the Killing form on $g$ by $\langle\> ,\>
\rangle$. The normalization is chosen such that long roots have square
length 2 and, hence, $S_{\rm WZW}[h]$ is uniquely defined modulo $2\pi
{\Bbb Z}$~[\Ref{FGK},\Ref{WITTEN}]. Finally, $\hbar m$ is a constant with
dimensions of mass, and $\beta$  is a coupling constant that has to be
quantized if the quantum theory is to be well defined; namely, $\hbar\beta^2
= 1/k$, where $k$ is a positive integer~[\Ref{WITTEN}]. The
Planck constant is explicitly shown to exhibit that, just as in the sine-Gordon
theory, the semi-classical limit is the same as the weak-coupling limit, and
that both are recovered when $k\rightarrow
\infty$. 

The action~\Act\ is invariant with respect to
the abelian gauge transformations
$$
h(x,t)\mapsto {\rm e\>}^\alpha\> h \> {\rm
e\>}^{-\>\tau(\alpha)}\>, \qquad A_\pm \mapsto A_\pm \>-
\>\partial_\pm\alpha\>,
\nfr{Gtrans}
where $\alpha=\alpha(x,t)$ takes values in $g_{0}^0 = u(1)^{+ r_g}$.
The precise form of the group of gauge transformations is specified by
$\tau$, which is an automorphism of $g_{0}^0$ that is compatible with the
Killing form of $g$. In our case, since $g_{0}^0$ is abelian and the restriction 
of
$\langle\> ,\> \rangle$ to $g_{0}^0$ is euclidean, $\tau$ is an element of the
group $O(r_g)$. However, in order to ensure that the resulting theory has a
mass-gap, $\tau$ has to be chosen as follows~[\Ref{MASS}]. Let us consider
the $x_\pm$-independent field configuration $h_0$ corresponding to the vacuum
of the theory or, equivalently, to the  absolute minimum of the potential.
The vacuum configuration $h_0$ satisfies $[\Lambda_+ , h_{0}^\dagger\>
\Lambda_-\> h_{0}]= 0$ and, since $\Lambda_\pm$ are regular, it induces, by
conjugation, an inner automorphism of $G$ that fixes the subalgebra
$g_{0}^0$, {\it i.e.},
$h_{0}^\dagger\> g_{0}^0\> h_{0} = g_{0}^0$. Then, the condition that the
theory has a mass-gap requires that the automorphism $\tau$ is chosen such
that 
$$
\tau(u)\not= h_{0}^\dagger\> u\> h_{0}
\nfr{NoTau}
for any $u$ in $g_{0}^0$~[\Ref{MASS}]. Otherwise, the choice of $\tau$ is
completely free and it provides supplementary continuous coupling
constants, besides $\Lambda_\pm$.

The resulting HSG theory exhibits an abelian $U(1)^{\times r_g}$ global
symmetry with respect to the group of  transformations~\note{Condition~\NoTau\
allows one to split any global transformation $h\mapsto {\rm e\>}^\alpha\> h
\> {\rm e\>}^\beta$ as the composition of a global gauge
transformation~\Gtrans\ and a global symmetry transformation in a unique way.}
$$
h(x,t)\mapsto {\rm e\>}^\alpha\> h \> {\rm
e\>}^{-\>(h_{0}^\dagger\>\alpha \> h_{0})}\>, 
\nfr{Global}
where $\alpha$ is a constant, $x_\pm$-independent, element of
$g_{0}^0$; these transformations fix the vacuum configuration~$h_0$. 
In order to give the expression of the corresponding Noether current,             
let us briefly describe the gauge-fixing procedure. It can be viewed as 
the choice of
some `canonical' form
$h^{\rm can}$ (gauge slice) such that any $h$ can be taken to that form 
through a
non-singular $h$-dependent gauge transformation,
$$
h\mapsto h^{\rm can} \>= \> {\rm e\>}^{\phi^{\rm can}[h]} \> h\>
{\rm e\>}^{ -\tau\left(\phi^{\rm can}[h]\right)} \>.
\nfr{GSlice}
Then, the gauge invariant conserved Noether current is~[\Ref{MASS}]
$$
J^\mu \> =\> \epsilon^{\mu\> \nu}\> \left(A_\nu \> +\> \partial_\nu
\phi^{\rm can}\right)\>.
\nfr{NoetCur}

In the following, we will assume that the vacuum configuration corresponds to
$h_0= I$ (up to gauge transformations), which can always be achieved through
a change of variables $h\mapsto h_0 h$. This implies that $\tau \not= I$ and,
hence, that the group of gauge transformations will not be of vector type
while global transformations~\Global\ will always be of vector type. 

Concerning gauge-fixing, we will use the `Leznov-Saveliev 
prescription' (LS)~[\Ref{MASS}]. It consists in choosing $A_\pm
=0$, which specifies the form of $h^{\rm can}$ only up to global gauge
transformations. This choice is viable due to the on-shell flatness of
the gauge field considered on two-dimensional Minkowski space. In the LS
gauge, the equations-of-motion reduce to 
$$
\eqalignno{
& \partial_-\left(h^{\dagger}\partial_+h\right)\> =\> -\> m^2
\left[\Lambda_+\> ,\> h^{\dagger} \Lambda_- h\right]\>, &
\nameali{Toda} \cr
& P\bigl(h^{\dagger}\partial_+h\bigr) \> = \>
P\bigl(\partial_-h\> h^{\dagger}\bigr) \> = \> 0\>, &
\nameali{Const}\cr}
$$
where $P$ is the projector onto the subalgebra $g_{0}^0$. Eq.~\Toda\ is
a non-abelian affine Toda equation~[\Ref{LS},\Ref{LUIZ},\Ref{NAT}],
and~\Const\ are the constraints that specify the form of the gauge
slice in this case. These constraints cannot be  solved
locally~[\Ref{PARK1}]. However, using the method of Leznov and
Saveliev, the explicit general solution of the non-abelian Toda
equation along with the constraints~\Const\ can be obtained in a
systematic way by means of the representation theory of affine
Kac-Moody algebras~[\Ref{LS}]. In particular, the multi-soliton
solutions are obtained, in this gauge, by means of the so-called
`solitonic specialization'~[\Ref{OLIVET},\Ref{OLIVE}] or, equivalently,
through the method of dressing transformations~[\Ref{TAUS}].

It is worth noticing that HSG theories do not exhibit parity invariance for
generic values of the coupling constants $\Lambda_\pm$ and $\tau$.
To be specific, parity invariance requires that
$$
\Lambda_+\> =\> \mu\> h_{0}^\dagger \Lambda_- h_{0} \quad
{\rm and}\quad \tau(u)\> =\> -\> h_{0}^\dagger\> u\> h_{0}\>,
\nfr{Parity}
for any $u$ in $g_{0}^0$, and some real number $\mu$~[\Ref{MASS}].
Then the parity transformation is $x\mapsto -x$, $h\mapsto h_0
h^\dagger h_0$, and $A_\pm \mapsto - A_\pm$.   

\chapter{Multi-Soliton solutions of HSG theories }

In this section, we work out the solitonic specialization of the
Leznov-Saveliev solution corresponding to the equations~\Toda\ and~\Const\
following the approach and conventions of~[\Ref{TAUS}], which is based on the
method of dressing transformations. First of all, we have to exhibit the
relationship of these equations with affine Kac-Moody algebras. 

Let $g_\cc$ be the complexification of the compact
Lie algebra $g$, and consider the central extension of its loop
algebra, $\gg$ (see the appendix). Following the approach
of~[\Ref{TAUS}], the eqs.~\Toda\ and~\Const\ are generalized non-abelian
Toda equations associated with the homogeneous gradation $\sh=(1,0,\ldots,0)$
of the untwisted affine Kac-Moody algebra $g_{\cc}^{(1)}= \gg \oplus {\Bbb
C}d$ and with the vacuum solution specified by $l=1$, $t_{\pm 1} = \pm
x_{\pm 1}$,
$$
{\cal A}_{1}^{({\rm vac})}\> =\> m \> \Lambda_{+}^{(1)}\>, \quad{\rm
and}\quad {\cal A}_{-1}^{({\rm vac})}\> =\> m \> \Lambda_{-}^{(-1)}\>+ \>
m^2\> \langle
\Lambda_{+}\>, \>\Lambda_{-}\rangle\> x_+\> c
\nfr{Data}
(see Section~IV.2 of~[\Ref{TAUS}], and recall that $[\Lambda_{+}^{(1)},
\Lambda_{-}^{(-1)}] = \langle \Lambda_{+}, \Lambda_{-}\rangle\> c$). Consider
a field $B=B(x_+, x_-)$ taking values in the finite Lie group formed by
exponentiating the subalgebra $\gg_0(\sh)$. Since $\gg_0(\sh) = g_{\cc}\oplus
{\Bbb C}\> c$, the field $B$ can be split as $B= h \> \exp(\nu \> c)$, where
$\nu=\nu(x_+, x_-)$ is the component along the central element $c$ and 
$h=h(x_+, x_-)$ takes values in $G_{\cc}$, the complex Lie group formed by
exponentiating $g_\cc$. Then, the approach of~[\Ref{TAUS}] provides a class
of solutions for the equation
$$
\partial_-\left(B^{-1}\partial_+B\right)\> =\> -\> m^2
\Bigl( \bigl[\Lambda_{+}^{(1)}\> ,\> B^{-1} \Lambda_{-}^{(-1)} B\bigr]\>
- \> \langle \Lambda_{+}\>, \>\Lambda_{-}\rangle\> c\Bigr) \>,
\nfr{TodaPlus}
subject to the constraints
$$
B^{-1}\partial_+B\> \in \> {\rm Im\/}\left({\rm ad}\>
\Lambda_{+}^{(1)}\right) \>, \quad{\rm and}\quad
\partial_-B \> B^{-1}\> \in \> {\rm Im\/}\left({\rm ad}\>
\Lambda_{-}^{(-1)}\right)\>.
\nfr{ContPlus}
Taking into account $B= h \> \exp(\nu \> c)$, eq.~\TodaPlus\ can be
split as
$$
\eqalignno{
& \partial_-\left(h^{-1}\partial_+h\right)\> =\> -\> m^2
\bigl[\Lambda_{+}\> ,\> h^{-1} \Lambda_{-} h\bigr] \>, &
\nameali{TodaPBack} \cr
& \partial_-\partial_+ \nu\> =\> -\> m^2 \> \langle \Lambda_{+}\> ,\>
h^{-1} \Lambda_{-} h \> -\> \Lambda_{-} \rangle\>, &
\nameali{TodaCent} \cr}
$$
while the constraints~\ContPlus\ become
$$
P_\cc(h^{-1}\partial_+h) \>= P_\cc(\partial_-h \> h^{-1})\> =\>0\>,
\nfr{ContPBack}
where $P_\cc$ is the projector onto the centralizer of $\Lambda_{\pm}$ in
$g_{\cc}$.

The resulting solutions are expressed by means of the integrable
highest-weight representations of $g_{\cc}^{(1)}$ (see the appendix for
the notation). Since these equations are related to the
homogeneous gradation~$\sh$, one needs the fundamental
representation $L(0)$, whose highest-weight vector is $\ket{v_0}$, to identify
the value of $\nu$, and another highest-weight representation $L(\tilde{\bf
s})$ such that the subset of vectors which are annihilated by all the
elements in $\gg_{>0}(\sh)$ form a faithful representation of $G_\cc$. Then,
these vectors can be labelled by means of the weights of the representation,
$\ket{\mub_0}, \ket{\mub_{0}'}$, and the solutions are given by~[\Ref{TAUS}]
$$
\eqalignno{
& \bra{\mub_{0}'} h^{-1}  \ket{\mub_0}\> =\> { \bra{\mub_{0}'} 
{\rm e\>}^{m\> x_+ \Lambda_{+}^{(1)}}\> {\rm e\>}^{-\>m\>
x_- \Lambda_{-}^{(-1)}}\> b\> 
{\rm e\>}^{m\> x_- \Lambda_{-}^{(-1)}} \> {\rm e\>}^{-\> m\> x_+
\Lambda_{+}^{(1)}} \ket{\mub_0} \over
\bra{v_0} {\rm e\>}^{m\>
x_+ \Lambda_{+}^{(1)}}\> {\rm e\>}^{-\>m\>
x_- \Lambda_{-}^{(-1)}}\> b\> 
{\rm e\>}^{m\> x_- \Lambda_{-}^{(-1)}} \> {\rm e\>}^{-\> m\> x_+
\Lambda_{+}^{(1)}} \ket{v_0}}\>, & \nameali{Solution}\cr 
\noalign{\vskip0.3cm}
& {\rm e\/}^{-\nu}\> = \> \bra{v_0} {\rm e\>}^{m\> x_+
\Lambda_{+}^{(1)}}\> {\rm e\>}^{-\>m\> x_- \Lambda_{-}^{(-1)}}\> b\> 
{\rm e\>}^{m\> x_- \Lambda_{-}^{(-1)}} \> {\rm e\>}^{-\> m\> x_+
\Lambda_{+}^{(1)}} \ket{v_0}\> \equiv \> \tau_0\>, &
\nameali{TauCero}\cr}
$$ 
for any constant element $b$ in the Kac-Moody group associated with
$g_{\cc}^{(1)}$.

The class of solutions summarized by eqs.~\TauCero\ and~\Solution\ are
conjectured to include the multi-soliton
solutions~[\Ref{OLIVET},\Ref{OLIVE},\Ref{TAUS},\Ref{LUIZ}]. They should
correspond to group elements which are products of exponentials of common
eigenvectors of $\Lambda_{+}^{(1)}$ and $\Lambda_{-}^{(-1)}$:
$$
b\> =\> {\rm e\/}^{F_1} \> {\rm e\/}^{F_2} \> \cdots \>, \qquad 
[ \Lambda_{\pm}^{(\pm1)} \> , \> F_k ] = \omega_{\pm}^{(k)} \> F_k \>.
\nfr{Eigenb}
In our case, and using the Chevalley basis introduced in the appendix, these
eigenvectors are associated with the roots of $g$. Namely, for each
root~$\alb$ and complex number~$z$,
$$
F_{\alb}(z)\> =\> \sum_{n\in {\Bbb Z}} z^{-n}\> E_{\alb}^{(n)}
\nfr{EigenV}
is an eigenvector of $\Lambda_{\pm}^{(\pm1)} = \pm\> i\lab_{\pm} \cdot
\hb^{(\pm1)}$ whose eigenvalues are
$$
\omega_{\pm}^{\alb}(z)\> =\> \pm\> i\> z^{\>\pm1} \>(\alb\cdot
\lab_{\pm})\>.
\nfr{EigenVal}
In these equations, $\lab_{\pm}$ are the components of $\Lambda_\pm$ in
that basis, and they have to be chosen in the same Weyl chamber of the Cartan
subalgebra to ensure that $h_0=I$ actually corresponds to the minimum of
$V(h)$. Recall that the fundamental particles are associated with the roots of
$g$, and the mass of the particle associated with $\alb$ is~[\Ref{MASS}]
$$
m_{\alb}\> =\> 2m\> \sqrt{(\alb\cdot \lab_{+})\> (\alb\cdot
\lab_{-})}\>.
\nfr{MassPart}

Eqs.~\TodaPBack\ and~\ContPBack\ are identical to the
equations-of-motion of the HSG theories in the LS gauge; namely,
eqs.~\Toda\ and~\Const, respectively. However, in the former~$h$
takes values in the complex Lie group $G_\cc$, while in the original
problem it has to take values in the compact Lie group $G$. Therefore,
it is necessary to restrict the choice of the arbitrary group
element~$b$ to ensure that the resulting solutions for~$h$ take values
in~$G$. Recall that the elements of~$g\subset g_\cc$ are characterized
by means of the anti-linear involutive automorphism~$\theta$ defined
by eq.~(A.2). In a similar fashion, the elements of the central extension of
$g_\cc$ which are compatible with the condition that
$\gg_{0}(\sh)\bigm|_{c=0}$ is compact can be singled out through the
following extension of~$\theta$ to~$\gg$:~\note{Notice that~$\hat\theta$ is
not the `Chevalley involution' of $g_{\cc}^{(1)}$, which is determined by
$\omega(E_{\alb_i}^{(0)}) = - E_{-\alb_i}^{(0)}$, for $i=1,\ldots, r_g$,
and $ \omega(E_{\alb_0}^{(+1)}) = - E_{-\alb_0}^{(-1)}$ and, hence,
$\omega\bigl(\gg_n(\sh) \bigr)= \gg_{-n}(\sh)$.}
$$
\hat\theta\Bigl(u^{(m)}\Bigr) \> =\> \Bigl(\theta(u)\Bigr)^{(m)}\>,
\quad \hat\theta(c)\>=\> c\>.
\nfr{Exten}
This automorphism of $\gg$ has the characteristic property of preserving the
homogeneous gradation, $\hat\theta \bigl(\gg_n(\sh) \bigr)= \gg_n(\sh)$, which
is needed because, in this case, homogeneous grades characterize
transformation properties with respect to the two-dimensional Lorentz group.
Then, the condition that products of the form~\Eigenb\ are fixed by
$\hat\theta$ requires that, instead of exponentials of eigenvectors, one has
to consider exponentials of combinations of the form
$$
a\> F_{\alb}(z)\> -\> a^\ast\> F_{\alb}^{\>\dagger} (z)\>,
\nfr{EigenComp}
for each root~$\alb$ and complex numbers~$a$ and~$z$. Since these
elements are not common eigenvectors of~$\Lambda_{\pm}^{(\pm1)}$, this
prescription can be somehow considered as a generalization of the original
solitonic specialization designed to ensure that the resulting solutions take
values in a compact group. Therefore, we conjecture that all the multi-soliton
solutions of eq.~\Toda, constrained by~\Const, can be obtained from
eq.~\Solution\ by considering group elements~$b$ which are products of
exponentials of elements of the form given by eq.~\EigenComp. 

In the LS gauge, the determination of the mass and charge carried by the
solitons is greatly simplified, and their relation with the boundary
conditions becomes explicit. Since the HSG theories exhibit a unique vacuum,
the boundary condition satisfied by soliton solutions at $x\rightarrow
\pm\infty$ is that their field configurations become $h_0$, up to gauge
transformations. However, using the LS gauge-fixing prescription $A_\pm=0$,
the only remnant gauge freedom corresponds to global transformations
and, hence, solitons satisfy
$$
\eqalign{
\lim_{ x\rightarrow \pm\infty} h(x,t) \> & =\>
{\rm e\>}^{q_\pm} \> h_0\>  {\rm e\>}^{-\tau(q_\pm)} \cr
&= \> {\rm e\>}^{q_\pm\> -\> h_0\tau(q_\pm)h_{0}^\dagger} \> h_0\> =
\> {\rm e\>}^{\xi_\pm} \> h_0\>, \cr} 
\nfr{Boundary}
where $q_\pm$ are constant elements in $g_{0}^0$. 

Let us first consider the calculation of the $U(1)^{\times r_g}$
charge. In the LS gauge, the conserved Noether current~\NoetCur\ simplifies to
$J^\mu = \epsilon^{\mu\>\nu} \partial_\nu \phi^{\rm can}$ and, hence, the
corresponding conserved charge is~\note{Our convention for the antisymmetric
tensor is $\epsilon_{0 1}= \epsilon^{1 0} = +1$.} 
$$
i\> {\bfmath Q}\cdot \hb \> =\> \int_{-\infty}^{+\infty} dx\> J^0\> =\>
\phi^{\rm can}[h( -\infty, t)] \>-\> \phi^{\rm can}[h(+\infty, t)] \>.
\efr
Then, taking into account the definition of $\phi^{\rm can}= \phi^{\rm
can}[h]$, eq.~\GSlice, and the boundary condition~\Boundary, the value of the
$U(1)^{\times r_g}$~vector charge is $q_- \>-\> q_+$. However, the
condition~\NoTau\ allows one to define the conserved charge as 
$$
i\> {\bfmath Q}\cdot \hb\> \equiv \> {1\over2 \pi \beta^2}\> (\xi_+ \>-\>
\xi_-)\>,
\nfr{SolChar}
which is easier to derive directly from the asymptotic behaviour of soliton
solutions in practice, and where the normalization has been chosen to
simplify the results of Section~6.

Next, let us address the calculation of the energy-momentum tensor,
which provides the mass of the solitons. Since the energy-momentum tensor
is gauge-invariant~[\Ref{MASS}], it can be calculated directly in the LS
gauge where the equations-of-motion of the HSG theories become
non-abelian affine Toda equations. Then, the calculation is largely
simplified by considering the results of~[\Ref{LUIZ}], where it is shown
that the energy-momentum tensor of the theory splits in two pieces
$$
\eqalignno{
& T_{\mu\> \nu}\> =\> \Theta_{\mu\> \nu} \> -\> {1\over 2\pi \beta^2}\>
\bigl(\partial_\mu \partial_\nu\> -\> g_{\mu\> \nu}\> \partial_\rho
\partial^\rho \bigr)\> \widehat{\nu}\>, \cr 
\noalign{\vskip0.1cm}
& {\rm where} \quad \widehat{\nu}\> =\> \nu\> -\>
m^2\> \langle \Lambda_+\>, \> \Lambda_-\rangle\> x_+ x_-\>. &
\nameali{SplitT}}
$$
This is the generalization to the non-abelian case of a similar
expression originally obtained in the context of abelian affine Toda
theories~[\Ref{OLIVET},\Ref{ARATYN}]. In~\SplitT, $\Theta_{\mu\> \nu}$ is
the traceless energy-momentum tensor of a related conformal affine
Toda theory associated with the Kac-Moody algebra $g_{\cc}^{(1)}$, which
vanishes for multi-soliton solutions~[\Ref{LUIZ},\Ref{OLIVET},\Ref{ARATYN}].
This means that the energy and momentum carried by solitons are given by the
second term in~\SplitT\ and, therefore, they arise through the boundary values
of~$\nu=\nu(x,t)$, the component of the field $B$ along the central element
of~$g_{\cc}^{(1)}$ (see eq.~\TodaPlus). Namely, 
$$
\eqalignno{
& P\> =\> \int_{-\infty}^{+\infty} dx\> T_{01}\> =\> {1\over 2\pi
\beta^2}\> \bigl(\partial_t \nu ( -\infty, t) \>-\> \partial_t
\nu(+\infty, t) \bigr)\>, & \nameali{Momentum} \cr
\noalign{\vskip0.1cm}
& E\> =\> \int_{-\infty}^{+\infty} dx\>
\bigl(T_{00}\> - {1\over
\beta^2} V(h_0) \bigr)\> =\> {1\over 2\pi
\beta^2}\> \bigl(\partial_x \nu ( -\infty, t) \>-\> \partial_x
\nu(+\infty, t) \bigr)\>, & \nameali{Energy} \cr}
$$
where the value of $V(h_0)/\beta^2$ has been subtracted to ensure
that the vacuum configuration has $E=P=0$.

\chapter{Soliton solutions of simply-laced HSG theories}

In this section we analyze the soliton solutions of the HSG theories
associated with simply-laced Lie algebras. In this case, the soliton
solutions can be expressed by means of the level-one fundamental
representations of~$g_{\cc}^{(1)}$ realized through the
homogeneous vertex operator construction~[\Ref{HOMVO},\Ref{GO},\Ref{BERN}]
(see the appendix). Therefore, in eq.~\EigenV, $F_{\alb}(z) =  V_{\alb}(z)$,
and the multi-soliton solutions are obtained from eq.~\Solution\ by
considering group elements $b$ which are products of elements of the
form~\note{Since we follow the conventions of~[\Ref{BERN}] for vertex
operators, notice that
$E_{\pm\alb}^{\>\dagger}= - E_{\mp\alb}$ and, hence, $F_{\alb}^\dagger(z) = 
-V_{-\alb}(z^\ast)$ (see the appendix).}
$$
\eqalign{
b_\alb (a,z)\> & =\> {\rm e\>}^{a\> V_{\alb}(z)\> +\> a^\ast \>
V_{-\alb}(z^\ast) }\> = \> \bigl( 1\> + \>  a\>
V_{\alb}(z)\bigr)\> \bigl( 1\> + \>  a^\ast \>
V_{-\alb}(z^\ast)\bigr) \cr
& =\> 1\> + \> a\> V_{\alb}(z)\>  + \>  a^\ast \>
V_{-\alb}(z^\ast) \> +\> |a|^2\>
V_{\alb}(z)\> V_{-\alb}(z^\ast)\>, \cr}
\nfr{BasicSol}
where we have used the widely known nilpotency of $V_{\alb}(z)$, eq.~(A.17).
Notice that, since $b_{-\alb} (a,z) = b_\alb (a^\ast , z^\ast)$ and
$$
a\> = \> |a|\> {\rm e\>}^{i\phi}\quad {\rm and}\quad z\> =\> {\rm
e\>}^{\rho\> +\> i\varphi}
\nfr{Angles}  
are generic complex numbers, we can assume that $\alb$ is always a
positive root and that $(\alb \cdot \lab_{\pm}) >0$ (recall that $\lab_+$ and
$\lab_-$ have to be chosen in the same Weyl chamber).
In eq.~\Solution, the only effect of the conjugation of $b_\alb
(a,z)$ with ${\rm e\>}^{m\> x_+ \Lambda_{+}^{(1)}}\> {\rm e\>}^{-\>m\>
x_- \Lambda_{-}^{(-1)}}$ is the change 
$$
a\> \mapsto \> d_{\alb}(z)\>  =\>  a\>  {\rm e\>}^{\Gamma_\alb (z)} 
\efr
where 
$$
\eqalign{
\Gamma_{\alb} (z)\> & =\> m\> x_+\>
\omega_{+}^\alb(z)\> -\>  m\> x_-\> \omega_{-}^\alb(z)\cr 
&=\> -m_\alb\> \sin(\varphi)\> {x\> +\> v_\alb\>
t\over \sqrt{1\> -\> v_{\alb}^2}}\> +\> 
i\> m_\alb\> \cos(\varphi)\> {t\> +\> v_\alb\>
x\over \sqrt{1\> -\> v_{\alb}^2}}\>,\cr}
\nfr{Vestido}
where $m_{\alb}$ is the mass of the fundamental particle associated with
$\alb$ and 
$$
v_\alb\> =\> \tanh\Bigl[ \rho \> + \> {1\over2}\> \ln {(\alb\cdot 
\lab_{+})\over (\alb\cdot
\lab_{-})}\Bigr]\>.
\nfr{Velocidad}

There is a one-soliton solution associated with each (positive) root $\alb$
of $g$ that is recovered with $b=b_{\alb} (a,z)$. Its explicit
form can be easily obtained from eqs.~\Solution, (A.10), and~(A.16):
$$
\eqalignno{
\bra{\mub_{0}'} h^\dagger  \ket{\mub_0}\> & =\>
{1\over \tau_0}\> \biggl(\Bigl[ 1\> +\> {|z|^2\> |a|^2\over 
|z-z^\ast|^2}\> {\rm e\>}^{\Gamma_{\alb}(z)\> +\>
\Gamma_{\alb}^\ast(z)}\> \left(z\over
z^\ast\right)^{\alb\cdot\mub_0}\Bigr]\> \delta_{\mub_0, \mub_{0}'}\cr
\noalign{\vskip0.3cm}
& + \> \epsilon(\alb,\mub_0 - v_{\tilde{\bf s}})\> z^{1\> +\> \alb\cdot
\mub_0}\> a\> {\rm e\>}^{\Gamma_\alb(z)}\> \delta_{\mub_0+\alb,
\mub_{0}'}\cr
\noalign{\vskip0.3cm}
& + \> \epsilon(-\alb,\mub_0 - v_{\tilde{\bf s}})\> (z^\ast)^{1\> - \>
\alb\cdot \mub_0}\> a^\ast\> {\rm e\>}^{\Gamma_{\alb}^\ast(z)}\>
\delta_{\mub_0, \mub_{0}'+\alb}\biggr) \>, & \nameali{SLSol}\cr}
$$
where $v_{\tilde{\bf s}}$ is the highest-weight of the representation
$L(\tilde{\bf s})$ involved in eq.~\Solution, and
$$
\eqalign{
\tau_0\> & =\> {\rm e\>}^{-\nu}\> \> = \>   
1\> +\> {|z|^2\> |a|^2\over |z-z^\ast|^2}\> {\rm
e\>}^{\Gamma_{\alb}(z)\> +\> \Gamma_{\alb}^\ast(z)}\cr
& =\> 1\> +\> {|a|^2 \over 4\> \sin^2 \varphi }\> \exp\bigl[
-2m_\alb\> \sin(\varphi)\> {x\> +\> v_\alb\>
t\over \sqrt{1\> -\> v_{\alb}^2}} \big]\>.}
\nfr{TauCeroSol}

In level-one representations of $g_{\cc}^{(1)}$ where $g$ is simply-laced,
all the weights $\mub_0$ satisfy $\alb \cdot \mub_0 = 0, \pm1
$~[\Ref{GO}]~\note{This property easily follows from $\|E_{\pm\alb}^{(-1)} 
\ket{\mub_0}\|^2 =\bra{\mub_0} E_{\pm\alb}^{\dagger (1)} E_{\pm\alb}^{(-1)}
\ket{\mub_0} \geq 0$ and $E_{\pm\alb}^{(1)} \ket{\mub_0} = 0$.} and, hence,
that $E_{\alb}^{(0)} \ket{\mub_0} \not=0$ if, and only if, $\alb\cdot \mub_0
= -1$. Then, the only non-vanishing matrix elements of $h^\dagger$ are
$$
\eqalign{
& \bra{\mub_0 + \alb} h^\dagger \ket{\mub_0}\> = \> -\> \epsilon(\alb,
\mub_0 -v_{\tilde{\bf s}})\> u_{\alb}^\ast (x,t)\>  \cr
& \bra{\mub_0} h^\dagger \ket{\mub_0 + \alb}\> = \> \epsilon(\alb,
\mub_0 -v_{\tilde{\bf s}})\> u_\alb (x,t)\>, \quad{\rm if} \quad
\alb\cdot
\mub_0=-1 \>, \cr}
\nfr{ExpOne} 
and
$$
\bra{\mub_0} h^\dagger \ket{\mub_0}\> = \> \sqrt{1\> -\>
|(\alb\cdot \mub_0) \> u_\alb (x,t)|^2}\;\; {\rm e\>}^{-\> i\> (\alb\cdot
\mub_0)\>\eta_\alb(x,t)}\>, \quad{\rm for\; all}\quad \mub_0\>,
\nfr{ExpTwo}
where
$$
\eqalignno{
u_\alb(x,t) \> &=\> -\> |\sin\varphi|\>{ {\rm e\>}^{-\>i\> \phi_0}\>
\exp \Bigl(-\> i \> m_\alb \cos\varphi \> {t\> +\> v_\alb(x\>- \>
x_0)\over \sqrt{1\> -\> v_{\alb}^2}}\Bigr) \over
\cosh \Bigl( m_\alb \> \sin\varphi
\> {x\> -\> x_0\> +\> v_\alb\> t \over \sqrt{1\> -\>
v_{\alb}^2}}\Bigr) }\>, \quad {\rm and} & \nameali{CSGsol}\cr
\noalign{\vskip0.3cm}
\eta_\alb(x,t)\> & = \> -\> \varphi\> +\> \arctan\Bigl[ \tan\varphi\>
\tanh\bigl(m_\alb \> \sin\varphi \> {x\> -\> x_0\> +\> v_\alb\> t 
\over \sqrt{1\> -\> v_{\alb}^2}}\bigr)\Bigr] \>. &
\nameali{CSGphase}\cr}
$$
In other words, $h$ takes values in the representation of the regular
$SU(2)$ subgroup of $G$ generated by $E_{\pm\alb}^{(0)}$ provided by the
vertex operator construction, an observation whose importance will become
clear in the next Section. This representation is formed by the set of vectors
$\ket{\mub_0}$ and $\ket{\mub_0 + \alb}$ with $\alb \cdot\mub_0 =-1$, which
is a direct sum of spin-$1/2$ irreducible factors. In the last equations
we have introduced
$$
x_0\>  =\> {\sqrt{1\> -\> v_{\alb}^2} \over m_\alb\> \sin\varphi}\>
\ln\Bigm|{a\over 2\> \sin\varphi}\Bigm|\quad {\rm and}\quad
\phi_0\> = \> \phi\> +\> {v_\alb\over \tan\varphi}\>  \ln\Bigm|{a\over
2\> \sin\varphi}\Bigm|\>,
\nfr{Constants}
which correspond to the centre of mass of the soliton and its orientation
in the internal $U(1)^{\times r_g}$ space.

Using eqs.~\Momentum\ and~\Energy, the energy and momentum carried by this
soliton can be easily calculated from the asymptotic behaviour of
$\tau_0$, eq.~\TauCeroSol:
$$
P\> =\> E \> v_\alb \>  = {1\over \pi\beta^2}\> {m_\alb\>
|\sin\varphi| \over \sqrt{1\>- \> v_{\alb}^2}}\> v_\alb \>.
\nfr{EnMom}
Therefore, the configuration given by eq.~\SLSol\ actually corresponds to 
a relativistic soliton with mass
$$
M_\alb(\varphi)\> =\> {1\over \pi\beta^2}\> m_\alb\> |\sin\varphi|  
\nfr{SolMass}
moving with velocity $v_\alb$; notice that $|v_\alb|$ is always $<1$. In
their rest frame, $v_\alb=0$, these field configurations are not static but
periodic time-dependent solutions that rotate in the internal 
$U(1)^{\times r_g}$ space with angular velocity
$$
\omega_\alb(\varphi)\> = \> m_\alb\> \cos \varphi\>. 
\nfr{AngVel}

Taking eq.~\SolChar\ into account, the charge carried by the soliton 
can be obtained from the asymptotic behaviour of its field configuration:
$$
h^\dagger(\pm\infty, t) \> =\>
\cases{I &, if $\>\pm \sin\varphi >0$ \cr   
{\rm e\>}^{2\>i \>\varphi\> \alb\cdot \hb}  &, if $\>\pm \sin\varphi
<0\> $ .\cr}
\nfr{Asymp}
Therefore, the conserved charged carried by the soliton is
$$
{\bfmath Q}_{\alb}(\varphi)\> =\> {1\over \pi\beta^2}\> {\rm
sign}[\> \sin\varphi \>]\>  \varphi \> \alb \quad {\rm mod} \quad {1\over
\beta^2} \> \Lab_{R}\>,
\nfr{Charge}
which is uniquely defined modulo $1/\beta^2$ times any element of
$\Lab_{R}$, the root lattice of $g$; an ambiguity that does not modify the
asymptotic behaviour given by~\Asymp\ (recall that $g$ is simply-laced).

We can summarize the physical meaning of the two complex parameters $a=|a|\>
{\rm e\>}^{i\phi}$ and $z = {\rm e\>}^{\rho +i\varphi}$ that label the soliton
solutions associated with a (positive) root $\alb$ as follows. Firstly,
eq.~\Constants\ shows that the only role of $|a|$ and $\phi$ is  to specify
the position of the centre of mass of the soliton and its orientation in  the
internal space, respectively. Actually, the only effect of a global symmetry
transformation $h \mapsto {\rm e\>}^{i\> \mub\cdot \hb}\>  h\> {\rm e}^{-i\>
\mub\cdot \hb}$ is the shift $\phi \mapsto \phi + \mub\cdot \alb$, which is
equivalent to a translation in  the internal space. Secondly, and according to
eq.~\Velocidad, $\rho$ is the rapidity of the soliton up to a
constant, and $\varphi$ specifies the angular velocity of the soliton
motion in the internal $U(1)^{\times r_g}$ space together with its mass and
charge. 

Finally, let us investigate the range of values of $\varphi$ leading to
inequivalent soliton solutions. The transformation $\varphi\mapsto -\varphi$
is equivalent to $\eta_\alb(x,t) \mapsto \eta_\alb(x,t) - 2\varphi$ plus a
change in $x_0$ and $\phi_0$ that can be trivially absorbed in
$a$. The change $\eta_\alb(x,t) \mapsto \eta_\alb(x,t) - 2\varphi$ can be
induced by means of the axial-like global transformation $h \mapsto {\rm
e\>}^{i\>\varphi\> \alb\cdot \hb}\>  h\> {\rm e}^{i\>\varphi\> \alb\cdot \hb}$
that, taking eq.~\NoTau\ into account, can always be split as the composition
of a global gauge transformation and a global symmetry transformation. This
implies that the field configurations corresponding to $\varphi$ and
$-\varphi$ actually correspond to the same soliton. Moreover, the vacuum
configuration $h_0=I$ is recovered with $\varphi=0,\> \pi$. Taking into
account all this and the fact that $\varphi$ is an angular variable, we
conclude that all the inequivalent soliton solutions are associated with the
values of $\varphi\not=0$ modulo $\pi$ or, equivalent, with $\varphi\in
(0,\pi)$.

\chapter{Soliton solutions of arbitrary HSG theories}

The soliton solutions of the HSG theories corresponding to non-simply laced
algebras can be explicitly studied by means of appropriated `foldings' of
the vertex operator construction considered in the previous section. However,
the properties of soliton solutions are better understood by noticing that
eq.~\CSGsol\ provides just the one-soliton solutions of the complex
sine-Gordon equation (CSG)~[\Ref{CSGSOLg},\Ref{CSGSOLm},\Ref{PARK2}]
$$
\partial_+ \partial_- u\> + \> u^\ast \>
{\partial_+ u\> \partial_- u^\ast \over 1 -
|u|^2}\> +\> {m_{\alb}^2 \over4} \> u\> \bigl(1 -
|u|^2 \bigr)\> =\> 0\>,
\nfr{CSGEq}
whose relation with the integrable deformation of the $SU(2)/U(1)$ coset
conformal field theory by the first thermal operator was originally pointed
out by Bakas~[\Ref{BAK}] (see also~[\Ref{PARK1}]). Actually, the CSG theory
is nothing else but the HSG associated with $SU(2)$, and we have already
indicated that the soliton solutions constructed in the previous section are
described by field configurations $h$ taking values in a regular $SU(2)$
subgroup of $G$ (see the comments below eq.~\CSGphase). Then,
using the fundamental representation of $SU(2)$, the field can be
parameterized as 
$$
h \> = \> \pmatrix{ {\rm e\>}^{i\> \eta}\> \sqrt{1 - |u|^2} & -\>
u^\ast \cr u &  {\rm e\>}^{-i\> \eta}\> \sqrt{1 - |u|^2}\cr }\>,
\nfr{CSGMat}
and eq.~\CSGEq\ is the equation-of-motion of~\Act\ with a particular gauge
fixing~[\Ref{BAK},\Ref{PARK1},\Ref{PARK2}].

In the original references, the CSG equation was associated with the
Lagrangian~[\Ref{CSGGEN},\Ref{CSGSOLg},\Ref{CSGSOLm}]
$$
{\cal L}\> =\> {|\partial_\mu u |^2 \over 1 - |u |^2
}\> - \> m_{\alb}^2 \> | u |^2\>,
\nfr{CSGLag}
whose relation with the action~\Act\ is
$$
S^{\eightpoint CSG}[h, A_\pm]\> =\> {1\over 4\pi\beta^2} \> \int \>
d^2 x\> {\cal L}\>\>
\nfr{LocalAct}
Again, this relation involves a particular choice of the gauge fixing
prescription such that the resulting form of the action is a local
functional of $u$. Eq.~\CSGLag\ allows one to calculate the mass and charge of
$SU(2)$ solitons directly by means of the Noether theorem:
$$
\eqalignno{
&{M^{\eightpoint CSG} \over \sqrt{1\> -\> v_{\alb}^2}}\>  =\> {1\over
4\pi\beta^2}
\> \int_{-\infty}^{+\infty} \> dx\> \Bigl({|\partial_t u|^2\> +\>
|\partial_x u|^2 \over 1 - |u |^2 }\> + \> m_{\alb}^2
\> | u |^2\Bigr) \>, \cr 
\noalign{\vskip0.3cm}
&q \>  =\> {i\over 4\pi\beta^2}\>
\int_{-\infty}^{+\infty} \> dx\> {u^\ast \>
\partial_t u \> -\> u\> \partial_t u^\ast
\over 1 - |u |^2 }\>. &
\nameali{CSGdir}\cr}
$$
Evaluated on the time-dependent soliton solutions~\CSGsol, this yields
eq.~\SolMass\ for the mass, while the charge is given by    
$$
\eqalign{
& q(\varphi)\>= \>q(-\varphi) \>=
\> (-1)^n\> q(\varphi+ n\pi)\quad {\rm and} \cr
& q(\varphi)\> =\> {1\over \pi
\beta^2}\> \Bigl[\varphi\> + \> {\pi\over2} \> \bigl( {\rm
sign}[{\pi\over2} - \varphi]\> -1\bigr) \Bigr]\quad {\rm for}\quad
\varphi\in (0,\pi)\>. \cr}    
\nfr{CSGcharge}
Although this might provide an unambiguous value for the charge carried
by the soliton, we will keep the definition of the conserved charge given
by eq.~\Charge\ and, hence,  
$$
{\bfmath Q}^{\eightpoint CSG}(\varphi)\cdot \hb\> =\> \bigl( q(\varphi)\;\;
{\rm mod} \;\; {1\over\beta^2} \bigr) \sigma_3 \>,
\nfr{CSGchargeP}
where $\sigma_1$, $\sigma_2$, and $\sigma_3$ are the Pauli matrices.
This definition of the charge emphasizes its interpretation as an angular
variable, which is natural in the approach of the previous sections (see
eq.~\NoetCur). In the following, we will also need the value of the action
corresponding to the one-soliton solutions~\CSGsol:
$$
S^{\eightpoint CSG}(\varphi)\> = \> {2\over \beta^2} \>\Bigl[ \bigm|
\pi\beta^2 \> q(\varphi)\bigm|\> -\> | \tan\varphi |
\Bigr] \>.
\nfr{CSGActVal}

Eq.~\CSGMat\ provides the general form of the field $h$ in the fundamental
representation of $SU(2)$ or, in the context of eqs.~\Solution, \ExpOne,
and~\ExpTwo, in the fundamental representation $L(\tilde{\bf s}) = L(1)$ of
$A_{1}^{(1)}$. Let us introduce a different but equivalent parameterization
of $h$ in the same representation
$$
h\> = \> {\rm e\>}^{i\>{\eta\over2}\>\sigma_3}\> {\rm e\>}^{i\>\psi
[\cos\theta\> \sigma_1\> +\> \sin\theta\> \sigma_2]}\> {\rm
e\>}^{i\>{\eta\over2}\> \sigma_3}\> = \> 
\pmatrix{ {\rm e\>}^{i\> \eta}\> \cos\psi & i\sin\psi\> {\rm e\>}^{-i\>
\theta}\cr
i\sin\psi\> {\rm e\>}^{i\> \theta}  &  {\rm e\>}^{-i\> \eta}\> \cos\psi
\cr}\>.
\nfr{MatSigma}
It can be generalized to any representation of $SU(2)$ by means of
$$
h\> =\> {\rm e\>}^{i\>{\eta\over2}\> J_0}\> {\rm e\>}^{i\>\psi
[\cos\theta\> (J_+ + J_-)\> -\> i\>  \sin\theta\> (J_+ - J_-)]}\> {\rm
e\>}^{i\>{\eta\over2}\> J_0}\>,
\nfr{RepGen}
where $J_0$ and $J_\pm$ are the Chevalley generators or $su(2)$, {\it
i.e.\/}, 
$$
[J_0, J_\pm]=\pm2 J_\pm \quad {\rm and} \quad [J_+, J_-]=J_0\>.
\efr

Considering eq.~\RepGen, the one-soliton solutions of the HSG theory
associated with an arbitrary simple Lie group $G$ can be
obtained in the following way. Let $\alb$ be a positive root of
$g$ and construct the CSG one-soliton field configuration described,
in the fundamental representation of $SU(2)$, by
$$
h_{\alb}^{\rm CSG}\> =\> \pmatrix{ {\rm e\>}^{i\> 
\eta_{\alb}}\> \sqrt{1 - |u_{\alb}|^2} & -\> u_{\alb}^\ast  \cr
u_{\alb} &  {\rm e\>}^{-i\>
\eta_{\alb}}\> \sqrt{1 - |u_{\alb}|^2}\cr }\>,
\efr
where $u_\alb$ and $\eta_\alb$ are given  by eqs.~\CSGsol\
and~\CSGphase, respectively. Next, let us consider the regular
embedding of $SU(2)$ in $G$ defined through
$$
\iota_\alb \>: \> (J_+,\> J_-,\> J_0) \in su(2)_{\cc} \longmapsto  
\bigl(E_\alb ,\> {\rm sign}[B_\alb]\> E_{-\alb}, \>{2\over
\alb^2}\> \alb\cdot \hb\bigr) \in g_{\cc}  \>.
\nfr{Embed}
Then, the one-soliton solution associated with $\alb$ is given by
$$
h_{\alb}^{g}=\widehat{\iota}_\alb \bigl(h_{\alb}^{\rm CSG}\bigr)
\in G \>,
\nfr{EmbedField}
where $\widehat{\iota}_\alb$ is the lift of $\iota_\alb$ to
$SU(2)$. Actually, it is easy to show that $h_{\alb}^{g}$ satisfies
eq.~\Toda. Since $h_{\alb}^{g}$ is in the embedded $SU(2)$ subgroup of
$G$, the left-hand-side of~\Toda\ is an element of the embedded $su(2)$
subalgebra of $g$, and only the components of $\Lambda_\pm$ along the
Cartan generator $\alb\cdot \hb$ contribute to the right-hand-side.
Then, $\Lambda_\pm$ can be decomposed as
$$
\Lambda_\pm\> =\> \pm\> i\> \lab_\pm \cdot \hb\> =\> 
\pm\> i\> {(\alb\cdot \lab_\pm) \over \alb^2}\> \alb\cdot \hb \> +\>
\cdots\>,
\efr
where the dots indicate components which are orthogonal to $\alb\cdot
\hb$. Therefore, eq.~\Toda\ reduces in this case to
$$
\partial_-\left(h_{\alb}^{g\> \dagger}\partial_+ h_{\alb}^{g}\right)\>
=\> -\> \left(m_{\alb}\over 2\right)^2
\Bigl[\bigl({1\over \alb^2} \alb\cdot\hb\bigr) \> ,\> h_{\alb}^{g\>
\dagger} \>\bigl({1\over \alb^2} \alb\cdot\hb\bigr)\>
h_{\alb}^{g}\Bigr]\>, 
\efr
which is just the embedding in $g$ of the CSG equation satisfied by the
$SU(2)$ field configuration $h_{\alb}^{\rm CSG}$. 

Recall that the Killing form of $g$ is normalized such that long roots have
square length $2$ and, hence, for non-simply laced algebras, the normalization
of the Killing form is different in the original and the embedded $su(2)$
algebras. The relative normalization is given by
$$
\langle \iota_\alb(\cdot)\> ,\> \iota_\alb(\cdot) \rangle_{g}
\> = \> {2\over \alb^2}\> \langle \cdot\> ,\>\cdot \rangle_{su(2)}\>,
\nfr{Normaliza}
as can be easily checked through the value of $\langle J_+ \>, \> J_-
\rangle$. 

Therefore, we conclude that the inequivalent solitons of the HSG theory
associated with an arbitrary simple Lie group $G$ are provided by the field
configurations $h_{\alb}^{g}=\widehat{\iota}_\alb (h_{\alb}^{\rm CSG})$
with $\varphi\in (0,\pi)$ and $\alb$ a (positive) root of $g$. 
Notice that, in the particular case of simply-laced groups, eqs.~\ExpOne\
and~\ExpTwo\ are just the matrix elements of $h_{\alb}^{\rm CSG}$ in the
homogeneous vertex operator representation. Then, taking into account that
the calculation of both the energy-momentum tensor and the action involves
the Killing form of $g$, the mass and action of the soliton corresponding to
$\varphi$ are  
$$
\eqalignno{
& M_\alb(\varphi)\> =\> {1\over \pi \beta^2}\>{2\over \alb^2} \>
m_\alb \> |\sin\varphi| & \nameali{MassGen}\cr 
\noalign{\vskip0.3cm}
& S_\alb(\varphi)\> =\> {2\over \beta^2}\>{2\over \alb^2} \>\Bigl[ 
\bigm| \pi\beta^2 \> q(\varphi)\bigm|\> -\>  
| \tan\varphi | \Bigr] \>.
&\nameali{ActCen}\cr}
$$
The value of the $U(1)^{\times r_g}$ conserved
charge carried by this soliton can be obtained by means of the
generalization of eq.~\Asymp,
$$
h_{\alb}^{\rm CSG}(\pm\infty, t) \> =\>
\cases{I \; \buildchar{\longmapsto}{\widehat{\iota}_\alb}{} \; I&, if  $\>\pm
\sin\varphi >0$
\cr 
{\rm e\>}^{-2\>i \>\varphi\> \sigma_3}\; 
\buildchar{\longmapsto}{\widehat{\iota}_\alb}{}\; {\rm e\>}^{-2\>i 
\>\varphi\> {2\over \alb^2}\> \alb\cdot\hb}&, if $\>\pm \sin\varphi <0\> $
,\cr}
\nfr{AsympG}
which leads to
$$
{\bfmath Q}_\alb(\varphi)\> =\> q(\varphi)\> {2\over \alb^2}\> \alb 
\quad {\rm mod}\quad {1\over \beta^2 } \> \Lab_{R}^\ast\>,
\nfr{ChargeGen}
where $q(\varphi)$ is specified by eq.~\CSGcharge, and $\Lab_{R}^\ast$ is
the co-root lattice of $g$, {\it i.e.\/}, the lattice generated by the
co-roots $\alb^\ast = (2/ \alb^2)\> \alb$. Recall that, in their rest frame,
solitons are given by time-dependent solutions that rotate in the internal
$U(1)^{\times r_g}$ space with angular velocity
$$
\omega_\alb(\varphi)\> = \> m_\alb\> \cos \varphi\>,
\efr
Then, each soliton of charge ${\bfmath Q}_\alb(\varphi)$ has a partner
of charge ${\bfmath Q}_\alb (\pi - \varphi)= - {\bfmath Q}_\alb(\varphi)
$ and the same mass that rotates with opposite angular velocity and, hence,
that can be identified as its anti-particle.

The construction of soliton solutions presented in this Section is very
similar to the construction of monopole solutions in theories with an adjoint
Higgs field in the Prasad-Sommerfeld limit, {\it e.g.\/}, $N=2$ or $N=4$
supersymmetric gauge theories. In the latter case, monopole solutions are
obtained by embeddings of the $SU(2)$ spherically symmetric 't~Hooft-Polyakov
monopole~[\Ref{MONOS},\Ref{MONOSUSY}]. In a similar way, in the HSG theories,
$\Lambda_\pm$ are analogous to the Higgs field vacuum expectation value and
the role of the 't~Hooft-Polyakov monopole is played by the $SU(2)$ CSG
soliton. 

In our construction we have only considered the regular embeddings of $SU(2)$
in $G$. However, a logical question is whether more general embeddings
provide additional soliton solutions. Let us consider an arbitrary embedding
$\iota: (J_\pm, J_0)\in su(2) \mapsto g$. Since the asymptotic behaviour of
the embedded CSG soliton $h^\iota=\widehat{\iota}(h_{\alb}^{\rm CSG})$ is 
$$
h^\iota(\pm\infty, t)\> =\>
\cases{I &, if $\>\pm \sin\varphi >0$ \cr 
{\rm e\>}^{-2\>i \>\varphi\> \iota(J_0)}&, if $\>\pm \sin\varphi <0\> $, \cr}
\efr
we have to restrict ourselves to those embeddings such that $\iota(J_0)$ is in
the Cartan subalgebra $g_{0}^0$ and, hence,
$$
\iota(J_0)\>= \> {\bfmath f}\cdot\hb\>, \quad \iota(J_+)\> =\> \sum_{\alb
\in \Delta}\> c_\alb\> E_\alb\> \quad{\rm and}\quad \iota(J_-)\> =\> {\rm
sign}[B_\alb]\>\sum_{\alb
\in \Delta}\> c_\alb\> E_{-\alb}\>,
\efr
where $\Delta$ is some set of positive roots of $g$. Then, $h^\iota$ can
provide a solution of~\Toda\ only if $[\Lambda_\pm, \iota(su(2)_\cc)] =
\iota(su(2)_\cc)$, which implies that the inner products $\alb\cdot
\lab_\pm$ are equal for all the roots $\alb \in \Delta$. However, $\Lambda_\pm$
are regular elements of $g_{0}^0$ and, therefore, all this implies that
$\alb \pm \beb$ cannot be a root of $g$ for any $\alb$ and $\beb$ in $\Delta$.
In other words, $h^\iota$ provide a soliton solution only if
$\iota(su(2))$ is the principal $su(2)$ subalgebra of some regular
$A_1\oplus \cdots \oplus A_1$ subalgebra of $g$~(for a nice
review about $su(2)$ embeddings see~[\Ref{EMBED}] and references therein)
and, in this case, $h^\iota$ is the product of the soliton solutions
associated with all the roots in $\Delta$:
$$
h^\iota\> =\> \prod_{\alb\in \Delta}\> \widehat{\iota}_\alb\bigl(
h_{\alb}^{\rm CSG} \bigr)\>.
\efr
Actually, if $\alb \pm \beb$ are not roots of $g$ it is straightforward to
check that the product of
$h_{\alb}^g = \widehat{\iota}_\alb\bigl( h_{\alb}^{\rm CSG} \bigr)$  
and  $h_{\beb}^g$ is another solution of~\Toda\ without constraining the
value of $\alb\cdot \lab_\pm$ and $\beb\cdot \lab_\pm$, but this new solution
has to be understood as a superposition of two non-interacting solitons.
Therefore, we conclude that the solutions constructed by means of non-regular
embeddings of $SU(2)$ in $G$ correspond to the superposition of a certain
number of non-interacting solitons and, hence, that they do not provide new
soliton solutions different from those given by eq.~\EmbedField.

\chapter{Semi-classical spectrum}

Having constructed the classical soliton solutions of the HSG theories,
we will now address their quantization. At rest, soliton solutions
provide explicit periodic time-dependent solutions and, therefore, we can
apply the Bohr-Sommerfeld quantization rule: $S_\alb(\varphi) +
M_\alb(\varphi)\> T_\alb(\varphi) = 2\pi\> n$, where $T_\alb(\varphi)=2\pi
/|\omega_\alb(\varphi)|$ is the period of the soliton solution at rest and
$n$ is a positive integer.  In our case, it is important to notice that the
action~\Act\ is multi-valued (modulo $2\pi/
\beta^2$) and, consequently, the coupling constant
$\beta$ has to be quantized~[\Ref{FGK},\Ref{WITTEN}]~\note{Quantum corrections
are expected to induce a finite renormalization of $k\rightarrow k_R= k +
{\eightpoint\it N}$, as it happens in the CSG theory~[\Ref{CSGSOLm}];
however, the difference between $k$ and $k_R$ is negligible in the
semi-classical $k\rightarrow\infty$ limit, and it will be ignored in
the following.}
$$
\beta^2 \> =\> {1\over k}\>, \quad k\in {\Bbb Z}^+\>.
\efr

Let us consider the semi-classical quantization of the soliton solutions
associated with a positive root $\alb$ of $g$. Then, the results of the
previous section lead to
$$
S_\alb(\varphi)\> +\> M_\alb(\varphi)\> T_\alb(\varphi)\> =
\> 2\pi\> {2\over \alb^2} \> |q(\varphi) | \>,
\nfr{BohrSom}
and the quantization rule becomes
$$
|q(\varphi) |\> =\> {\alb^2 \over 2} \> n \>.
\efr
Classically, eq.~\CSGcharge\ implies that $0< |q(\varphi) |\leq k/2$.
Moreover, the value of $q(\varphi)$ for the time independent solution
obtained by setting $\varphi=\pi/2$ is not uniquely defined
$$
\lim_{\varphi \rightarrow {\pi\over2}^{\pm}} q(\varphi)\> =\> \mp\>
{k\over2}\>,
\efr
a property whose importance has been emphasized by Dorey and Hollowood in
relation to the CSG theory~[\Ref{TN}]. This means that $q(\varphi) = \pm
k/2$ actually represent the same classical soliton solution. This ambiguity
can be avoided by identifying the values of $q(\varphi)$ modulo $
k$~$(=1/\beta^2)$~[\Ref{TN}], an identification that is naturally
incorporated in our definition of the conserved charge ${\bfmath
Q}_\alb(\varphi)$, eq.~\ChargeGen.  
 
Taking into account all this, the semi-classical quantization of the
soliton solutions associated with a positive root
$\alb$ gives rise to $2k/\alb^2 - 1$ massive particles that can be
labelled by an integer number
$$
1\> \leq \>n \> \leq \> {2 k\over \alb^2}\> -1\>.
\efr
This number corresponds to $2 q(\varphi)/\alb^2$ or $2 (q(\varphi) +
k)/\alb^2$, depending on whether $q(\varphi)>0$ or $q(\varphi)<0$,
respectively. Then, the mass and charge carried by these particles
are
$$
\eqalignno{
M_\alb (n) \> & =\> {2 k\over \pi \alb^2}\> m_\alb \> \Bigm| \sin 
\Bigr( {\pi \alb^2 \over 2 k}\> n \Bigr)\Bigm| & \nameali{MassBS}\cr
\noalign{\vskip0.3cm}
{\bfmath Q}_\alb(n) \> &= \>  n\> \alb \quad {\rm mod}\quad k \>
\Lab_{R}^\ast\>, \quad n\>= \> 1,2,\ldots, {2k\over \alb^2}\> -\> 1\>, &
\nameali{ChargeBS} \cr}
$$
and, now, the anti-particle of the particle $n$ is labelled by
$2k/\alb^2- n$. 

Notice that the $U(1)^{\times r_g}$ conserved charge, although continuous in
the classical theory, should now more properly be though of as a discrete
charge that takes values in $\Lab_R$ modulo $k\> \Lab_{R}^\ast$. Actually, it
can be easily checked that the co-root lattice $\Lab_{R}^\ast$ is just the
lattice spanned by the long roots of $g$. Therefore, ${\bfmath Q}_\alb(n)$
takes values in the global symmetry group of the theory of $G$-parafermions at
level-$k$ described by the gauged WZW action $S_{\rm WZW}[h, A_\pm]$ in
eq.~\Act~[\Ref{GPARAF}], which points to the parafermionic character of the
solitonic spectrum. We will comment about this in the last Section.

As it could have been anticipated, for a fixed root $\alb$ the spectrum is
similar to the spectrum of the complex sine-Gordon
theory~[\Ref{CSGSOLm},\Ref{TN}], which itself resembles the spectrum of
breather states in the ordinary sine-Gordon theory. However, notice that the
number of particles resulting from the semi-classical quantization of the
solitons associated with a root $\alb$ depends on its length. Thus, for long
roots the number of resulting particles is always $k-1$, while for the short
roots of $B_n$, $C_n$ or $ F_4$ is $2k-1$, and for those of $G_2$ it is
$3k-1$. The lightest states in the spectrum of particles associated with a
given root correspond to $n=1$ and $n=2k/\alb^2 -1$, and they have charge 
${\bfmath Q}_\alb(1) = -\>{\bfmath Q}_\alb(2k/ \alb^2 -1) =  \alb$ and mass
$M_\alb(1) = m_\alb + O(1/k^2)$. These quantum numbers are identical to those
of the fundamental particle of the theory associated with the root $\alb$
(together with its anti-particle) and, therefore, we assume that these states
describe the fundamental particle in the semi-classical limit. Moreover,
${\bfmath Q}_\alb(n) = n \> {\bfmath Q}_\alb(1)$ and
$$
{ M_\alb (n) \over M_\alb (1)}\> =\> n \> -\> {\pi^2 \alb^4 \over
24 k^2} \> n\> (n^2\> -\> 1)\> +\> O(k^{-4})\>,
\nfr{Bound}
which suggests the obvious interpretation of the state labelled by $n$ as 
a bound-state of $n$ fundamental particles. Recall that all these
identifications are possible because solitons do not carry topological
charges, as it also happens in the CSG theory, and hence there is no
topological distinction between the vacuum sector and the one-soliton
sector.

A very peculiar property of these theories is that they exhibit unstable
fundamental particles and solitons. To spell this out, let us start with the
fundamental particles associated with three roots $\alb$, $\beb$, and
$\alb+\beb$. Then, it is easy to show that
$$
\eqalign{
&m_{\alb+\beb}^2\> =\> \bigl( m_{\alb}+ m_{\beb}\bigr)^2 \> +\>
4\> m_\alb\> m_\beb \> \sinh^2 \bigl(\sigma_\alb -
\sigma_\beb \bigr) \cr
& {\rm where}\quad \sigma_\alb\> =\> {1\over2}\> \ln {(\alb\cdot
\lab_{+})\over (\alb\cdot \lab_{-})}\>, \cr}
\efr
which implies that $m_{\alb+\beb}\geq m_{\alb}+ m_{\beb}$. Using this
relation, one can deduce a lower bound for the mass of the fundamental
particle associated with a root $\alb = \sum_{i=1}^{r_g} p_i \alb_i$, 
$$
m_\alb \geq \sum_{i=1}^{r_g} p_i\> m_{\alb_i}\>,
\nfr{DecayF}
where $\alb_1, \ldots, \alb_{r_g}$ is a set of simple roots. To understand
the implications of this bound for the soliton mass spectrum, let us
consider 
$$
\eqalign{
\sum_{i=1}^{r_g}\> M_{\alb_i}(n\> p_i)\> & =\>  
\sum_{i=1}^{r_g}\> {2 k\over \pi \alb_{i}^2}\> m_{\alb_i}\> \Bigm| \sin 
\Bigr( {\pi \alb_{i}^2 \over 2 k}\> n\> p_i \Bigr)\Bigm| \cr
& =\>  
{2 k\over \pi \alb^2}\> \sum_{i=1}^{r_g}\> {\alb^2\over \alb_{i}^2} \>
m_{\alb_i}\> \Bigm| \sin 
\Bigr( p_i\> {\alb_{i}^2 \over \alb^2}\>\> { \pi \alb^2 \over 2 k}\> n\>
\Bigr)\Bigm| \>,\cr}
\nfr{SolRel}
for an arbitrary positive integer number~$n$. Notice that $p_i =
(2/\alb_{i}^2)\> \lab_i\cdot \alb$, where $\lab_i$ are the fundamental
weights of $g$ ($\alb_i\cdot \lab_j = (\alb_{i}^2 /2)\> \delta_{i,j}$) and,
hence, $p_i\>(\alb_{i}^2 /\alb^2) = \lab_i \cdot \alb^\ast$ is a positive
integer number. Then, using eq.~\DecayF\ and the relation $|\sin (m\theta)|
\leq m |\sin (\theta)|$ that is satisfied for $m\geq1$, eq.~\SolRel\ becomes
$$
\eqalign{
\sum_{i=1}^{r_g}\> M_{\alb_i}(n\> p_i)\> & \leq \>  
{2 k\over \pi \alb^2}\> \Bigl(\sum_{i=1}^{r_g}\> p_i\>
m_{\alb_i}\Bigr)\> \Bigm| \sin 
\Bigr( { \pi \alb^2 \over 2 k}\> n\> \Bigr)\Bigm| \cr
&\leq {2 k\over \pi \alb^2}\> m_\alb \> \Bigm| \sin 
\Bigr( { \pi \alb^2 \over 2 k}\> n\> \Bigr)\Bigm| \> = \> M_\alb(n)\>.\cr}
\nfr{SolDecay}
This bound, together with
$$
{\bfmath Q}_{\alb}(n) \> =\> \sum_{i=1}^{r_g} {\bfmath Q}_{\alb_i}(n\>
p_i) \>,
\efr
shows that either the soliton particle labelled by $(\alb, n)$ is unstable and
decays into solitons associated with the simple roots or, if the
bound~\SolDecay\ is saturated, it is a bound-state at threshold. Similar
phenomena occur for monopoles and dyons in $N=2$ and $N=4$ supersymmetric
gauge theories~[\Ref{MONOSUSY}]. 

Recall that unstable particles do not correspond to asymptotic scattering
states and, hence, the only trace of them should be the existence of
resonance poles in the $S$-matrix of the theory. As a consequence of the
previous analysis, the scattering states of the HSG theories are expected to
be described only by the solitons associated with the simple roots. These
peculiarities will make the determination of the exact $S$-matrix of these
theories and its confrontation with the results of semi-classical
quantization extremely interesting. 

\chapter{Conclusions.}

In this paper we have obtained the semi-classical spectrum of the
HSG theories associated with arbitrary compact simple Lie groups. At the
quantum level, these theories are massive perturbations of the level-$k$
$G$-parafermions~[\Ref{GPARAF}], where $1/k$ is the (quantized) coupling
constant of the theory. There is a classical one-soliton solution associated
with each root $\alb$ of $g$, which can be constructed by embedding a CSG
soliton in the $SU(2)$ subgroup of $G$ generated by $E_{\alb}$, $E_{-\alb}$,
and $[E_{\alb}, E_{-\alb}]$. For each root, the resulting semi-classical
spectrum consists of a tower of $2k/\alb^2 -1$ massive particles. The
lightest one can be identified with the fundamental particle associated with
the root $\alb$, which means that the full spectrum of the HSG theory is
described by solitons. This supports the expectation that it should be
possible to infer the form of the exact $S$-matrix by means of semi-classical
methods, and makes possible to compare semi-classical results with standard
perturbation theory in the large-$k$ limit.  

Unlike the kinks of the sine-Gordon equation, the solitons of the HSG theories
do not have topological quantum numbers, but they carry a conserved vector
Noether charge ${\bfmath Q}$ associated with a global $U(1)^{r_g}$ symmetry. 
After quantization, the charge is restricted to take values in $\Lab_R$ modulo
$k\Lab_{R}^\ast$, {\it i.e.\/}, the root lattice of $g$, modulo $k$ times the
co-root lattice. Therefore, since $\Lab_{R}^\ast$ equals the long root
lattice, the quantized charge carried by the solitons actually takes values
in the discrete symmetry group of Gepner's theory of $G$-parafermions at
level~$k$~[\Ref{GPARAF}]. 

The possibility of a semi-classical solitonic interpretation of parafermions
was originally suggested by Bardakci {\it et al.\/}~[\Ref{BCR}], but the
relation between their results and ours is unclear. However, it is interesting
to speculate on the identification of the HSG solitons with parafermionic
fields. Following Gepner~[\Ref{GPARAF}], and denoting by $\Lab_W$ the weight
lattice of $g$, each field in the theory of level-$k$ $G$-parafermions has a
charge $ (\lab, \overline{\lab})$ taking values in $\Lab_W \times \Lab_W$
modulo $k \Lab_{R}^\ast \times k \Lab_{R}^\ast$, and constrained by $\lab -
\overline{\lab} \in \Lab_R$. On the other hand, for $h_0=I$, the global
symmetry transformations of the HSG theories, eq.~\Global, are of vector
type, which makes the $U(1)^{r_g}$ charge carried by the $(\lab,
\overline{\lab})$ field to be precisely ${\bfmath Q}= \lab - \overline{\lab}$.
Therefore, the soliton particle labelled by $(\alb, n)$ should be related to
fields whose charge is of the form $(n\alb + \lab, \lab)$. In particular, the
fundamental particle (anti-particle) associated with the root $\alb$ could well
correspond to the `generating parafermions' $\psi_\alb(z)$ ($\overline{
\psi}_\alb( \overline{z})$), although we do not have any definite argument to
support a conjecture in that sense. 

An important result is that some of the soliton particles are unstable, which
means that the HSG theories actually capture important features of
monopoles and dyons in $N=2$ and $N=4$ supersymmetric gauge theories in four
dimensions~[\Ref{MONOSUSY}]. To be precise, for $r_g\geq2$, only the solitons
associated with simple roots give rise to stable particles while the other
ones are either unstable or bound-states at threshold. Therefore, in the 
$S$-matrix formalism, only stable soliton particles are expected to provide
asymptotic scattering states, while the existence of unstable particles
should manifest as resonance poles. A better understanding of these features
requires a careful study of the scattering of solitons which will be
presented elsewhere.

Recently, an integrable perturbation of the $su(2)_k$ WZW model that also
exhibits stable and unstable particles has been constructed by
Brazhnikov~[\Ref{BRA}]. Actually, one can show that this model is included in
the clasification of~[\Ref{MASS}] as a SSSG theory associated with the
symmetric space $SU(3)/SO(3)$. This way, it can be generalized to provide a
new series of integrable perturbation of WZW models that share most of the
properties of the model constructed by Brazhnikov; in particular, the presence
of stable and unstable particles in the spectrum~[\Ref{SPLIT}]. These novel
features make the HSG and SSSG theories quite unconventional because, up to 
our knowledge, the bootstrap program has been so far applied only to theories
whose spectrum consists entirely of stable particles. All this provides extra
motivation for the overall aim of finding their factorizable $S$-matrix.

\bjump\jump
\centerline{\bf Acknowledgements}

\jump
\noindent
We would like to thank T.J.~Hollowood for his early collaboration in this
work, and J~S\'anchez Guill\'en, L.A.~Ferreira, and F.~Ravanini for many useful
discussions. We also thank M.V.~Gallas for his contribution to some of the
calculations in this paper. This research is supported partially by CICYT
(AEN96-1673) and DGICYT (PB93-0344). We are also grateful for partial support
from the EC Comission via a TMR Grant, contract number FMRX-CT96-0012, and
from NATO (CRG~950751).

\bjump 
   
\appendix{Appendix: Conventions about Lie and Kac-Moody algebras}

Given a compact simple Lie algebra~$g$ and Lie group~$G$, we shall use
a subscript `$\cc$' to denote their complexifications, {\it i.e.},
$g_\cc$ and $G_\cc$, respectively. Let us introduce a Chevalley basis
for $g_\cc$ which consists in Cartan generators $\mub\cdot\hb$ together
with step operators $E_\alb$ for each root $\alb$, such that $\mub$
and $\alb$ live in a $r_g={\rm rank}(g)$-dimensional vector
space provided with an inner product normalized such that long roots have
square length~$2$. These obey
$$
[\mub\cdot\hb\>, \> E_\alb]\>=\> (\mub\cdot \alb)\> E_\alb\>, \quad
{\rm and}\quad
[E_\alb\>, \> E_{-\alb}]\> =\> B_\alb\> \alb\cdot\hb\>,
\nfr{CWbasis}
and the basis is completely specified by choosing $B_\alb=\langle
E_\alb, E_{-\alb} \rangle$ to be either $+ 2/ \alb^2$ or $- 2/
\alb^2$, and the sign is fixed once for all the roots. The first choice is
standard in the context of matrix realizations of $g$, while the second
is very convenient when considering vertex operator constructions.

Given~$g_\cc$, its compact real form~$g$ can be characterized through an
automorphism $\theta$ of $g_\cc$ that is `anti-linear', meaning $\theta(a_1
\varphi_1 + a_2 \varphi_2) = a_{1}^\ast\> \theta(\varphi_1) +  a_{2}^\ast\>
\theta(\varphi_2)$ for any
$a_1, a_2\in {\Bbb C}$ and $\varphi_1, \varphi_2\in g_\cc$, and
`involutive', meaning $\theta^2=I$. The relevant automorphism is defined
through
$$
\theta(E_{\pm\alb})\> =\> -\>
{\rm sign}[B_\alb]\> E_{\mp\alb} \equiv\> -\> E_{\pm\alb
\>}^\dagger  \>, \quad 
\theta(\hb)\>=\> -\>\hb\> \equiv\> -\>\hb^\dagger\>.
\nfr{AutoC}
Then, $g$ consists of those elements of $g_\cc$ fixed by the
automorphism: $\theta(\varphi)=\varphi$.

The central extension of the loop algebra associated with $g_\cc$ can be
written as
$$
\eqalignno{
&\gg= \bigl\{ u^{(m)} \>\bigm|\> u\in g_\cc\>,\; m \in {\Bbb Z}
\bigr\}\> \oplus \> {\Bbb C}\> c\>,  
\cr
\noalign{\vskip 0.2cm}
&\bigl[ u^{(m)}\> ,\> v^{(n)}\bigr] \>= \> [u\>
,\>v]^{(m+n)}\> +
\> m\>  \langle u\>,\> v\rangle \> c\> \delta_{m+n,0}\>, \cr
&\bigl[ d\>, \> u^{(m)}\bigr] \>= \> m\> u^{(m)}\>, \qquad \bigl[ c\>,
\> d\bigr] \> =\> \bigl[ c\>, \> u^{(m)}\bigr]\> =\> 0\>. &
\nameali{Loop} \cr}
$$
Here, ${\Bbb C} c$ is the centre of~$\gg$, and~$d$ is the derivation
that induces the `homogeneous gradation',
$$
\eqalignno{
& \gg \> =\> \bigoplus_{n\in {\Bbb Z}} \gg_n(\sh)\>, \cr
& \gg_n(\sh)\> = \> \bigl\{ u^{(n)} \>\bigm|\> u\in g_\cc
\bigr\}\> \oplus\> \bigl\{{\Bbb C}\> c \bigr\} \> \delta_{n,0}\>, &
\nameali{HomGrad}\cr}
$$
which can be labelled by the vector $\sh=(1,0,\ldots,0)$.~\note{Recall
that the different integer gradations of~$\gg$ are labelled by sets
$\ss=(s_0,\ldots,s_{r_g})$ of non-negative integers that specify the
grades of the Chevalley generators $e_{i}^{+}$~[\Ref{TAUS},\Ref{KBOOK}].}
This way, $g_\cc^{(1)}\> =\> \gg \oplus {\Bbb C}\>d$ is an `untwisted
affine Kac-Moody algebra'~[\Ref{KBOOK}] whose Chevalley generators are
$$
e_{i}^+ \>=\> \cases{E_{\alb_i}^{(0)}\>, & for 
$i=1,\ldots,r_g\>$,\cr
\noalign{\vskip 0.3cm}
E_{\alb_0}^{(1)}\>, & for $i=0\>$,\cr}\qquad
h_i \>=\> {2\over \alb_{i}^2}\> \alb_i\cdot \hb^{(0)} \> +\> c\>
\delta_{i,0}\>.
\nfr{LoopGen}
In these equations, $\alb_1, \ldots, \alb_{r_g}$ are simple
roots of $g_\cc$, and $-\alb_0 = \sum_{i=1}^{r_g} k_i\> \alb_i$ is the
highest root, which is always a long root and, hence, $\alb_{0}^2 =2$.

Multi-soliton solutions are constructed by means of the `integrable'
highest-weight representations of~$g_\cc^{(1)}$, which can also be
labelled by
gradations~$\ss=(s_0,\ldots,s_{r_g})$~[\Ref{TAUS},\Ref{KBOOK}]. The
highest-weight vector $\ket{v_\ss}$ of the representation $L(\ss)$
satisfies
$$
e_{i}^+\> \ket{v_\ss} \>= \> \left(e_{i}^-\right)^{s_i +1} \> \ket{v_\ss}
\>=\>0 \>,  \quad h_i \> \ket{v_\ss} \>= \> s_i \> \ket{v_\ss} \>,
\efr
for all $i=0,\ldots,r$. The eigenvalue of the centre $c$ on $L(\ss)$
is known as the level $k$, 
$$
c\> \ket{v_\ss} \>= \> \sum_{i=0}^r k_{i}^{\vee}\> h_i \> \ket{v_\ss} 
\>= \> \left(\sum_{i=0}^r k_{i}^{\vee}\> s_i\right) \>\ket{v_\ss}\>,
\nfr{Centre}
where $k_{i}^{\vee}=(\alb_{i}^2/2)\> k_i$ are the labels of the dual
Dynkin diagram of $g_{\cc}^{(1)}$. Therefore, in these representations,
$k= \sum_{i=0}^r k_{i}^{\vee}\> s_i$, and the level is always a
positive integer. Following~[\Ref{TAUS}], we will use the notation
$L(i)$ and $\ket{v_i}$ for the fundamental representation and
highest-weight vector corresponding to $\ss={\bf s}^{(i)}$ with
$s^{(i)}_j=
\delta_{i,j}$, whose level equals $k_{i}^{\vee}$. Then, the
highest-weight vector of
$L(\ss)$ can be decomposed as
$$
\ket{v_\ss} \>=\> \bigotimes_{i=0}^{r}\> \bigl\{ \ket{v_i}^{\otimes s_i}
\bigr\}\>.
\nfr{Tensor}

For simply laced affine Kac-Moody algebras, all the fundamental integrable
representations of level-one are isomorphic  to the `basic representation'
$L(0)$, which can be realized by means of vertex operators acting on Fock
spaces~[\Ref{VO},\Ref{HOMVO},\Ref{GO},\Ref{BERN}]. Then, the other fundamental
integrable representations of level~$>1$ can be realized as submodules in the
tensor product of several fundamental level-one representations. Moreover,
the fundamental integrable representations of non-simply laced Kac-Moody
algebras can be constructed from those of the simply laced algebras by folding
them~[\Ref{TUROLB},\Ref{MARCO}].

In Section~4 the soliton solutions of the HSG theory associated with a
simply-laced Lie algebra $g$ are constructed by means of the
(level-one) `homogeneous vertex operator
construction'~[\Ref{HOMVO},\Ref{GO},\Ref{BERN}], which can be summarized as
follows. First of all, if the representation is of level-one, the generators
of the form $\mub\cdot \hb^{(n)}$ satisfy the (homogeneous) Heisenberg
algebra 
$$
[\mub\cdot \hb^{(m)}\>, \> \gab\cdot \hb^{(n)}]\> =\> m\> (\mub\cdot
\gab)\> \delta_{m+n,0}\>, 
\efr   
which is equivalent to the commutation relations of $r_g$ free bosonic
fields. Thus, they can be represented as a set of oscillators
$\hb^{(n)}\mapsto {\bfmath a}_n$ acting on a Fock space. Let $\ket{0}$ be
the vacuum vector for these oscillators and ${\cal F}$ the Fock space
representation of the Heisenberg algebra spanned by the oscillators and the
identity operator. Next, we identify ${\bfmath a}_0= {\bfmath p}$ as a
momentum operator and introduce a corresponding conjugate position operator
${\bfmath q}$ such that
$$
[{\bfmath q}^j\>, \> {\bfmath p}^k] \> =\> i\> \delta^{j,k}\>,
$$
and it commutes with all the other oscillators. Let $W$ be the
infinite-dimensional vector space spanned by vectors of the form
$\ket{\alb} = {\rm e\>}^{i\> \alb\cdot {\bfmath q}}\ket{0}$ where $\alb$ is
in the root lattice $\Lab_R$ of $g$; obviously,  ${\bfmath p}\ket{\alb}=
\alb \ket{\alb}$. For any root $\alb$ of $g$, let us define the vertex
operator
$$
\eqalign{
V_\alb(z) \> &= \sum_{n\in {\Bbb Z}} z^{-n}\> V_{\alb}^n
\cr &=\> z\> \exp \Bigl[ \sum_{n>0} {z^n\over n}\> \alb\cdot
{\bfmath a}_{-n}\Bigr]\> \exp \Bigl[ -\> \sum_{n>0} {z^{-n}\over n}\>
\alb\cdot {\bfmath a}_{n}\Bigr]\> {\rm e\>}^{i \> \alb\cdot {\bfmath q}}\>
z^{\> \alb\cdot {\bfmath p}}\> C_\alb\>.\cr}  
\nfr{VOdef}
In this definition, the $C_\alb$'s are a set of operators known as
`Klein factors' or `cocycle operators', which act as
$$
C_\alb \> \ket{\beb}\> =\> \epsilon(\alb, \beb) \ket{\beb}\>, \quad {\rm
for\;\; all}\quad \beb\in \Lab_R\>,
\efr
where $\epsilon(\alb, \beb)$ equals $\pm1$ and satisfies certain
consistency conditions. Concerning $\epsilon(\alb,
\beb)$, we will follow the conventions of ref.~[\Ref{BERN}], which,
in particular, imply that $\epsilon(\alb, \beb)$ is a 2-cocycle and that
$\epsilon(\alb, -\alb)= \epsilon(\alb, \alb)=-1$. 

Then, the homogenous vertex operator construction provides a realization of
the basic representation $L(0)$ on the Hilbert space ${\cal F}\otimes W$ as
follows:
$$
\hb^{(n)} \longmapsto {\bfmath a}_n\>, \qquad E_{\alb}^{(n)}\longmapsto
V_{\alb}^n\>.
\efr
It is important to point out that, in this representation and with the
conventions of~[\Ref{BERN}], the generators
${E}_{\alb}^n$ satisfy the commutation relations
$$
[{E}_{\alb}^n\> , \> {E}_{\beb}^m]\> =\> \cases{
\epsilon(\alb, \beb)\> {E}_{\alb +\beb}^{n+m} & if
$\alb+\beb$ is a root of $g$ \cr
-\> \alb \cdot {\bfmath a}_{n+m}\> -\> m\> \delta_{m+n,0} & if
$\alb+\beb=0\>, $\cr}
\efr
which implies the choice of $B_\alb =-1$ in eq.~\CWbasis. The homogenous
vertex operator construction also provides a realization of the other
fundamental integrable representations of level-one. Let us consider a
fundamental weight $\lab_i$ of $g$, {\it i.e.\/}, $\lab_i\cdot
\alb_j =\delta_{i,j}$, such that $k_{i}^\vee =1$ (recall that $g$ is
simply-laced). Then the fundamental representation $L(i)$ is realized through
the same construction based on a different vacuum 
$$
\ket{0}\longmapsto \ket{\lab_i}\> =\> {\rm e\>}^{i\> {\bfmath q}\cdot
\lab_i}\> \ket{0} \>,
\efr
together with a trivial modification of the Klein factors:
$$
C_\alb \> \ket{\lab_i + \beb}\> =\> \epsilon(\alb, \beb)\> \ket{\lab_i +
\beb}\>,
\quad {\rm for\;\; all}\quad \beb\in \Lab_R\>.
\efr

The product of two vertex operators can be normal-ordered such that
$$
\eqalign{
V_\alb(z)\> & V_\beb(w) \> =\> z\> w\> (z-w)^{\alb\cdot \beb}\>
\exp \Bigl[ \sum_{n>0} {z^n \alb + w^n \beb \over n}\> \cdot
{\bfmath a}_{-n}\Bigr]\cr
& \exp \Bigl[ -\> \sum_{n>0} {z^{-n} \alb + w^{-n} \beb \over n}\>
\cdot {\bfmath a}_{n}\Bigr]\>
\epsilon(\alb, \beb)\>  {\rm e\>}^{i \> (\alb +\beb)\cdot {\bfmath q}} \>
C_{\alb+\beb}\>  z^{\> \alb\cdot {\bfmath p}}\> w^{\> \beb\cdot {\bfmath
p}} \>,\cr} 
\nfr{TwoVO} 
which implies the nilpotency of the vertex operator,
$$
V_\alb(z)\> V_\alb(z) \> =0\>,
\nfr{Nilpot}
and the relation $ V_\alb(z) V_\beb(w) = V_\alb(w) V_\beb(z)$ for $z\not=w$. 

\bjump
\references

\beginref
\Rref{CSGSOLg}{B.S.~Getmanov, JETP Lett. {\bf 25} (1977) 119; \newline
I.V.~Barashenkov and B.S.~Getmanov, Commun. Math. Phys. {\bf 112} (1987)
423.}
\Rref{BCR}{K.~Bardakci, M.~Crescimanno and E.~Ravinovici, Nucl. Phys. {\bf B
344} (1990) 344.}
\Rref{SG}{R.F. Dashen, B.~Hasslacher, and A.~Neveu, Phys. Rev. {\bf D
11} (1975) 3424; \newline
A.B. Zamolodochikov and Al. B. Zamolodchikov, Ann. Phys. {\bf120} (1979)
253.}
\Rref{SKYR}{T.H.R.~Skyrme, Proc. R.~Soc. London {\bf A 260} (1961) 127.}
\Rref{ADK}{G.~Adkins, C.~Nappi and E.~Witten, Nucl. Phys. {\bf B 228} (1983)
552.}
\Rref{MODY}{G.~'t~Hooft, Nucl. Phys. {\bf B 79} (1974) 276; \newline
A.M.~Polyakov, JETP Lett. {\bf 20} (1974) 194; \newline
B.~Julia and A.~Zee, Phys. Rev. {\bf D 11} (1975) 2227; \newline
E.~Tomboulis and G.~Woo, Nucl. Phys. {\bf B 107} (1976) 221.}
\Rref{PARAF}{A.B.~Zamolodchikov and V.A.~Fateev, Sov. Phys. JETP
{\bf 62} (1985) 215; \newline
D.~Gepner and Z.~Qiu, Nucl. Phys. {\bf B 285} (1987) 423.}
\Rref{GPARAF}{D.~Gepner, Nucl. Phys. {\bf B 290} (1987) 10; \newline
G.~Dunne, I.~Halliday and P.~Suranyi, Nucl. Phys.~{\bf B 325} (1989) 526.}
\Rref{EMBED}{L.~Frappat, E.~Ragoucy and P.~Sorba, Commun. Math. Phys. {\bf
157} (1993) 499.}
\Rref{MONOS}{F.A.~Bais, Phys. Rev.~{\bf D 18} (1978) 1206;\newline
E.J.~Weinberg, Nucl. Phys.~{\bf B 167} (1980) 500; Nucl. Phys.~{\bf B 203}
(1982) 445.}
\Rref{MONOSUSY}{J.P. Gauntlett and D. A. Lowe, Nucl. Phys. {\bf B
472} (1996) 194;\newline 
K. Lee, E.J. Weinberg and P. Yi, Phys. Lett. {\bf B 376} (1996) 97; Phys.
Rev. {\bf D 54} (1996) 1633; \newline 
O.~Aharony and S.~Yankielowicz, Nucl. Phys.~{\bf B 473} (1996) 93;\newline 
T.J.~Hollowood, {\sl Semi-classical decay of monopoles in $N=2$ gauge
theory\/}, hep-th/9611106; \newline
C.~Fraser and T.J.~Hollowood, Nucl. Phys.~{\bf B 490} (1997) 217; {\sl
Semi-classical quantization in $N=4$ supersymmetric Yang-Mills theory and
Duality\/}, hep-th/9704011.}
\Rref{INST}{C.W.~Bernard, N.H.~Christ, A.H.~Guth and E.J.~Weinberg, Phys.
Rev. {\bf D 16} (1977) 2967; \newline
C.W.~Bernard, Phys. Rev. {\bf D 19} (1979) 3013;\newline
S.F.~Cordes, Nucl. Phys. {\bf B 273} (1986) 629.}
\Rref{BRA}{V.A.~Brazhnikov, {\sl $\Phi^{(2)}$ Perturbations of WZW
Model\/}, Rutgers Univ. prep. RU~96-110, hep-th/9612040.}
\Rref{CSGSOLm}{H.J.~de Vega and J.M.~Maillet, Phys. Rev. {\bf D 28} (1983)
1441.}
\Rref{CSGGEN}{K.~Pohlmeyer, Commun. Math. Phys. {\bf 46} (1976) 207;
\newline F.~Lund and T.~Regge, Phys. Rev. {\bf D 14} (1976) 1524;
\newline F.~Lund, Phys. Rev. Lett. {\bf 38} (1977) 1175.}
\Rref{ARATYN}{H.~Aratyn, L.A.~Ferreira, J.F.~Gomes, and A.H.~Zimerman,
Nucl. Phys.~{\bf B 406} (1993) 727.}
\Rref{GO}{P.~Goddard and D.~Olive, Int. J.~Mod. Phys. {\bf A 1} (1986)
303.}
\Rref{KBOOK}{V.G.~Kac, {\sl Infinite
dimensional Lie algebras\/} ($3^{rd}$ ed.), Cambridge
University Press (1990).} 
\Rref{MASS}{C.R.~Fern\'andez-Pousa, M.V.~Gallas, T.J.~Hollowood, and
J.L.~Miramontes, Nucl. Phys.~{\bf B 484} (1997) 609.}
\Rref{SPLIT}{C.R.~Fern\'andez-Pousa and J.L.~Miramontes, work in
preparation.}
\Rref{QNOS}{C.R.~Fern\'andez-Pousa, M.V.~Gallas, T.J.~Hollowood, and
J.L.~Miramontes, {\sl Solitonic Integrable Perturbations of
Parafermionic Theories\/}, hep-th/9701109, to appear in Nucl. Phys.~{\bf
B}.}
\Rref{WITTEN}{E.~Witten, Commun. Math. Phys. {\bf 92} (1984) 455.}
\Rref{NAT}{J.~Underwood, {\sl Aspects of Non-Abelian Toda Theories\/},
Imperial/TP/92-93/30, hep-th/9304156.}
\Rref{LUIZ}{L.A.~Ferreira, J.L.~Miramontes, and J.~S\'anchez
Guill\'en, Nucl. Phys. {\bf B 449} (1995) 631.}
\Rref{PARK1}{Q-H.~Park, Phys. Lett. {\bf B 328} (1994) 329.}
\Rref{PARK2}{Q-H.~Park and H.J.~Shin, Phys. Lett. {\bf B 359} (1995)
125.}
\Rref{LS}{A.N.~Leznov and M.V.~Saveliev, Commun. Math. Phys. {\bf
89} (1983) 59; {\sl Group theoretical methods for integration of
non-linear dynamical systems\/}, Prog. Phys.~15 (Birkhauser, Basel,
1992).}
\Rref{OLIVET}{D.I.~Olive, N.~Turok and J.W.R.~Underwood, Nucl. Phys.
{\bf B 401} (1993) 663.}
\Rref{OLIVE}{D.I.~Olive, N.~Turok and J.W.R.~Underwood, Nucl. Phys. {\bf
B 408} (1993) 565; \newline D.I.~Olive, M.V.~Saveliev and
J.W.R.~Underwood, Phys. Lett. {\bf B 311} (1993) 117.}
\Rref{TAUS}{L.A.~Ferreira, J.L.~Miramontes and J.~S\'anchez Guill\'en,
J.~of Math. Phys. {\bf 38} (1997) 882.}
\Rref{TUROLB}{D.~Olive, N.~Turok and J.W.R.~Underwood, Nucl. Phys.  {\bf
B409} (1993) 509.}
\Rref{MARCO}{A.~Fring, P.R.~Johnson, M.A.C.~Kneipp, and D.I.~Olive, Nucl. 
Phys. {\bf B430} (1994) 597; \newline
M.A.C.~Kneipp, and D.I.~Olive,  Commun. Math. Phys. {\bf 177} (1996) 
561.}
\Rref{VO}{J.~Lepowsky and R.L.~Wilson, Commun. Math. Phys. {\bf 62} (1978)
43; \newline V.G.~Kac, D.A.~Kazhdan, J.~Lepowsky and R.L.~Wilson, Adv. in
Math. {\bf 42} (1981) 83;\newline
J.~Lepowsky, Proc. Natl. Acad. Sci. USA {\bf 82} (1985) 8295; \newline
V.G.~Kac and D.H.~Peterson, {\sl 112 constructions of
the basic representation of the loop group of $E_8$\/}, in `Symposium
on Anomalies, Geometry and Topology' (W.A.~Bardeen and
A.R.~White, eds.), World Scientific (1985) 276.}
\Rref{HOMVO}{I.B.~Frenkel and V.G.~Kac, Invent. Math. {\bf 62} (1980) 28;
\newline
G.~Segal, Comm. Math. Phys. {\bf 80} (1981) 301.}
\Rref{BERN}{D.~Bernard and J.~Thierry-Mieg, Comm. Math. Phys.
{\bf 111} (1987) 181.}
\Rref{FGK}{G.~Felder, K.~Gawedzki and A.~Kupianen, Comm. Math. Phys.
{\bf 117} (1988) 127.}
\Rref{BAK}{I.~Bakas, Int. J.~Mod. Phys. {\bf A 9} (1994) 3443.}
\Rref{TN}{N.~Dorey and T.J.~Hollowood, Nucl. Phys. {\bf B 440} (1995)
215.}
\Rref{ZAMINT}{A.B.~Zamolodchikov, Adv. Stud. Pure Math. {\bf 19} (1989)
641; Int. J. Mod. Phys. {\bf A3} (1988) 743; JETP Lett. {\bf 46}
(1987) 160.}
\Rref{EYHM}{T.~Eguchi and S-K.~Yang, Phys. Lett. {\bf B 224} (1989) 373;
\newline T.J.~Hollowood and P.~Mansfield, Phys. Lett. {\bf B 226} (1989) 73.}
\Rref{SOLC}{T.J.~Hollowood, Nucl. Phys. {\bf B~384} (1992) 523; \newline
H.~Aratyn, C.P.~Constantinidis, L.A.~Ferreira, J.F.~Gomes,
and A.H.~Zimerman, Nucl. Phys. {\bf B~406} (1993) 727.}
\Rref{JAPOS}{S.P.~Khastgir and R.~Sasaki, Prog. Theor. Phys. {\bf 95} (1996)
485.}

\endref   

\ciao